\documentclass[]{emulateapj}

\usepackage{hyperref} 
\usepackage{amsmath}

\usepackage[]{units} 
\usepackage{multirow}
\usepackage{soul}

\begin{document}
\title{Following the Cosmic Evolution of Pristine Gas I: \\ Implications for Milky Way Halo Stars}
\shorttitle{Evolution of Pristine Gas I}
\author{Richard Sarmento\altaffilmark{1,3}, Evan Scannapieco\altaffilmark{1,3}, \& Liubin Pan\altaffilmark{2}}
\altaffiltext{1}{School of Earth and Space Exploration,  Arizona State University, P.O. Box 871404, Tempe, AZ, 85287-1404}
\altaffiltext{2}{Harvard-Smithsonian Center for Astrophysics, 60 Garden St., Cambridge, MA 02138, USA}
\altaffiltext{3}{Joint Institute for Nuclear Astrophysics, USA}
\begin{abstract}

We make use of a new subgrid model of turbulent mixing to accurately follow the cosmological evolution of the first stars, the mixing of their supernova (SN) ejecta, and the impact on the chemical composition of the Galactic Halo. Using the cosmological adaptive mesh refinement code \textsc{ramses}, we implement a model for the pollution of pristine gas as described in Pan et al. Tracking the metallicity of Pop III stars with metallicities below a critical value allows us to account for the fraction of $Z < Z_{\rm crit}$ stars formed even in regions in which the gas's average metallicity is well above $Z_{\rm crit}.$  We demonstrate that such partially-mixed regions account for $0.5$ to $0.7$ of all Pop III stars formed up to $z=5$. Additionally, we track the creation and transport of ``primordial metals'' (PM) generated by Pop III SNe. 
These neutron-capture deficient metals are taken up by second-generation stars and likely lead to unique abundance signatures characteristic of carbon-enhanced, metal-poor (CEMP-no) stars.   As an illustrative example, we associate primordial metals with abundance ratios used by Keller et al. to explain the source of metals in the star SMSS J031300.36-670839.3, finding good agreement with the observed [Fe/H], [C/H], [O/H], and [Mg/Ca] ratios in CEMP-no Milky Way halo stars.  Similar future simulations will aid in further constraining the properties of Pop III stars using CEMP observations, as well as improve predictions of the spatial distribution of Pop III stars, as will be explored by the next generation of ground- and space-based telescopes.
\end{abstract}


\section{Introduction}

The death of the first stars resulted in the highly nonuniform pollution of the universe.   By this time big bang nucleosynthesis had produced helium efficiently, but it had been halted by the expansion of the universe before it could go much further, leaving stars and supernovae (SNe) to form and disseminate the heavier elements  \citep{1991ApJ...376...51W}. These early SNe first enriched the gas in and around protogalaxies which, in turn, led to a gradual, spatially inhomogeneous transition from metal-free Population III (Pop III) star formation to metal-enriched Population II (Pop II) star formation \citep{2003ApJ...589...35S,2006ApJ...653..285S,2007ApJ...661...10B,2007MNRAS.382..945T,2007ApJ...654...66O,2010ApJ...712..435T,2010MNRAS.407.1003M,2012ApJ...745...50W,2013ApJ...773..108C,2013MNRAS.428.1857J,2013ApJ...775..111P,2014MNRAS.440.2498P}.

Most of the properties of this transition remain unknown.  High-redshift observations have yielded candidates for Pop III stellar populations \citep{2002ApJ...565L..71M,2004ApJ...617..707D,2006Natur.440..501J,2007MNRAS.379.1589D,2008ApJ...680..100N,2012ApJ...761...85K, 2013A&A...556A..68C} including a single  $z=6.6$ galaxy with no detected metal emission lines and  narrow HeII $\lambda$1640 emission \citep{2015ApJ...808..139S} that is possibly indicative of metal-free stars \citep[e.g][]{2001ApJ...550L...1T,2002A&A...382...28S}.   However, these measurements are only able to hint at the overall progression of early metal enrichment.

Similarly, searches for individual Pop III stars in the Milky Way (MW) halo have yielded only indirect clues as to the Pop III/Pop II transition.  Even the most pristine stars observed have a substantial mass fraction of metals   \citep{2002Natur.419..904C, 2004A&A...416.1117C,2006ApJ...639..897A,2005Natur.434..871F,2007ApJ...670..774N,2011Natur.477...67C,2014Natur.506..463K,2015Natur.527..484H}, even though theoretical studies suggest that metal-free stars  should have already been observed if they had masses low enough to survive to the present-day \citep{2006ApJ...653..285S,2006ApJ...641....1T,2007ApJ...661...10B,2010MNRAS.401L...5S,2015MNRAS.447.3892H,2016arXiv160200465I}.  On the other hand, low-temperature cooling is extremely inefficient  without dust and metals, and pristine gas would have been much less susceptible to fragmentation, forming individual $\approx10^3 M_\odot$ stars, \citep{1976ApJ...205..103H, 2002Sci...295...93A, 2002ApJ...564...23B, 2003Natur.425..812B} or single or binary $10-100 M_\odot$ stars with very short lifetimes \citep[e.g.][]{2006MNRAS.366..247J, 2007ApJ...661..972P, 2008ApJ...681..771M, 2009Sci...325..601T, 2010MNRAS.403...45S, 2011ApJ...727..110C, 2011Sci...334.1250H, 2012MNRAS.424..399G,2014ApJ...781...60H}.

Theoretically, the key process determining the Pop III/Pop II transition is the turbulent mixing of heavy elements to pollute the medium above the critical metallicity, $Z_{\rm crit}$, that marks the shift to lower-mass star formation.  This evolution is dependent on two important issues. The first is the uncertain value of $Z_{\rm crit}$,  which is expected to be between $10^{-6}$  and $10^{-3} Z_{\odot},$ depending on whether low-metallicity cooling is dominated by dust emission or by the fine-structure lines of carbon and oxygen \citep{2003Natur.422..869S, 2003Natur.425..812B, 2005ApJ...626..627O}.  The second issue is the rate at which the gas within the galaxy can be polluted above $Z_{\rm crit}$ by the turbulent mixing of heavy elements \citep{2007ApJ...654L..29P,2010ApJ...716..510G,2012JFM...700..459P,2013ApJ...775..111P, 2013IAUS..295....3B, 2014MmSAI..85..202B, 2015MNRAS.451.1190R}.  Here the key quantity is the evolution of the pristine gas fraction,  the fraction of material with metal concentration below $Z_{\rm crit}$ as a function of time and space.  Clearly the  behavior of this quantity depends both on the rate at which stars deposit new metals into the surrounding medium and also on how gravity and large-scale motions act to move metals between different regions within that medium. However,  the evolution of the pristine gas fraction is also highly dependent on a third process: the small-scale turbulent mixing of metals within a given region of initially pristine gas.

This last process is particularly difficult to model in cosmological simulations, due to the enormous range of physical scales required to capture it properly.
Estimating the diffusivity of the ionized medium as the Spitzer-Braginskii value \citep{1956pfig.book.....S} 
\begin{equation}\label{eq:spitzer}
\begin{aligned}
\nu_{\rm spitzer} = \frac{6.0\times 10^{-17} (T/{\rm K})^{5/2} }{(\rho /{\rm g}/{\rm cm^{-3}})} {\rm cm}^{2} {\rm s^{-1}},
\end{aligned}
\end{equation}
gives $L_{\rm diss} \approx \nu_{\rm spitzer} /c_s \approx 10^{-3}\, T_{4}^{2}\, n_{-3}^{-1}\, {\rm pc}$ where $T_4$ is temperature in units of $10^4$ K and $n_{-3}$ is the number density in units of $10^{-3}$ cm$^{-3},$ which is the mean density at $z \approx 15.$ 
On the other hand, the maximum comoving resolution of modern large-scale cosmological simulations is $\approx 10-1000$ pc \citep[e.g.][]{2013ApJ...771...81R,2014ApJS..210...14K,2014MNRAS.444.1518V,2015ApJ...807L..12O,2015MNRAS.446..521S, 2014MNRAS.444.1453D}.

Given the mismatch between these spatial scales, thorough mixing of pollutants at the scale of individual resolution elements is expected to take several eddy turnover times, \citep{2013ApJ...775..111P,2015MNRAS.451.1190R} corresponding to a significant number of  simulation time steps.   However, most simulations instantaneously update the affected cells' average metallicity once they are contaminated with SN ejecta.  In other words, because such cells may have a relatively high average metallicity immediately after a local SN event, they are usually assumed to be fully polluted to above the critical metallicity, even though in reality their mass fraction, $P$, of unpolluted, pristine gas remains large until these metals are well mixed.   

In this work, we develop and apply a subgrid model that explicitly tracks the pristine gas fraction, $P$, within every zone in a cosmological simulation.  Our model is built on standard techniques for estimating unresolved turbulent velocities and the results of both theoretical modeling and high-resolution simulations of the pollution of pristine gas in fully developed turbulence, as described in \citet[hereafter PS13]{2013ApJ...775..111P} \citep[see also][]{2012JFM...700..459P}. The model is implemented within the cosmological adaptive mesh refinement (AMR) code \textsc{ramses} \citep{2002A&A...385..337T} which also includes models for star formation and metal generation/dispersal by SNe. 
 All quantities that include an overbar indicate cell or star-particle-averaged qualities. Hence, the average metallicity of a simulation cell is denoted $\overline Z$ while quantities including the \textit{star} subscript (e.g., $\overline Z_{\star}$) refer to values associated with simulated star particles. We also add another scalar quantity, $\overline Z_{\rm P}$, to the code ,which tracks the metallicity due only to metal-free stars (here, the subscript P is short for `primordial metals').  When star particles are formed within a given cell, they inherit not only the total metallicity $\overline Z$, but also $P$ and $\overline Z_{\rm P}$, from the gas.  This allows us to calculate the fraction of stars in a given star particle that are metal-free, $P_\star$, as well as the relative contributions that metals from Pop III and Pop II stars make to the stars that are enriched, $\overline Z_{\rm P,\star} / \overline Z_{\star}.$  

These values remained locked into the stellar populations for all times, and can be compared with observations of MW Halo stars with very low metallicities. We directly compare our simulation results with observations of the  metallicity distribution function of MW halo stars, and show that accounting for $P_\star$ is essential for making such comparisons reliably.  In addition, knowledge of $\overline Z_{\rm P,\star} / \overline Z_{\star}$ allows us to investigate the idea that stars formed in gas enriched only by Pop III stars are likely to be carbon-enhanced and especially iron-poor \citep{2003Natur.422..871U, 2014Natur.506..463K, 2016A&A...586A.160H}.

The structure of this work is as follows.  In \S2 we describe our methods, including both our implementation of the subgrid model of metal pollution and our \textsc{ramses} modeling of the overall evolution of cosmological objects and the formation of star particles within them. In \S3 we describe our results, focusing on comparisons between the properties of our final stellar distributions and observations of metal-poor stars in the MW Halo. Conclusions are given in \S4. Throughout this paper, we adopt the following cosmological parameters: $\Omega_{\rm M} = 0.267$, $\Omega_{\Lambda} = 0.733$, $\Omega_{\rm b} = 0.0449$, $h = 0.71$, $\sigma_8 = 0.801$, and $n = 0.96,$  where $\Omega_{\rm M}$, $\Omega_{\Lambda}$, $\Omega_{\rm b}$ are the total matter, vacuum, and baryonic densities, in units of the critical density, $h$ is the Hubble constant in units of 100 km/s, $\sigma_8$ is the variance of linear fluctuations on the 8 $h^{-1}$ Mpc scale, and $n$ is the ``tilt" of the primordial power spectrum \citep{2011ApJS..192...16L}.

\section{Methods}

\subsection{Primordial Metallicity and Pristine Gas Fraction}\label{sec:pmpg}

Our study makes use of \textsc{ramses} \citep{2002A&A...385..337T}, a cosmological AMR code, which uses an unsplit second-order Godunov scheme for evolving the Euler equations. \textsc{ramses} 
variables are cell centered and interpolated to the cell faces for flux calculations, which are then used by a Harten-Lax-van Leer-Contact Riemann solver ~\citep{1979JCoPh..32..101V,1988SJNA...25..294E}. Self-gravity is solved using the multigrid method \citep{GT2011} for all coarse levels in the simulation, and the conjugate gradient method is used for levels $\geq 11$. The code is capable of advecting any number of scalar quantities, defined as mass fractions, in each simulation cell.  For example, the standard version of \textsc{ramses} evolves a mass fraction of metals for each cell, which may have contributions from Pop III stars and second-generation stars, referred to as total average metallicity $\overline Z$. As a simulation evolves, \textsc{ramses} creates star particles in regions of overdense gas, each of which represents hundreds to thousands of solar masses of individual stars. Like the gas, each such star particle is tagged with a $\overline Z_{\star}$ value, representing the average metallicity of the medium from which it was born.  

We use these capabilities to generate and track new metallicity-related quantities for both the gas and star particles.  For the gas, we introduce two new scalars: the average \textit{primordial metallicity}, $\overline Z_{\rm P}$ and the \textit{pristine gas mass fraction}, $P$. The primordial metallicity scalar, $\overline Z_{\rm P}$, tracks the mass fraction of metals generated by Pop III stars, which are likely to have nonsolar abundance ratios \citep{2002ApJ...567..532H, 2003Natur.422..871U, 2014ApJ...792L..32I}.  Our pristine gas fraction scalar models the mass fraction of gas with $ Z < Z_{\rm crit}$ which allows us to track the process of metal mixing within each cell. While a cell may have a relatively high average metallicity, $\overline Z$, those metals are normally not well mixed throughout 
its volume at the time of injection. By tracking $P$, we can quantify the amount of pristine gas in such cells ($P \approx 1.0$) even if their mean metallicities are large.   

Since we are primarily interested in the characteristics of stellar populations, we also track the pristine fraction and the primordial metallicity of each star particle ($P_{\star}$ and $\overline Z_{\rm P,\star}$ respectively). 
These values are adopted for each star particle from the gas in which it was formed. For example, a star particle born in a region of gas with $P = 0.5$ inherits $P_{\star} = 0.5$ and represents a stellar population containing $50\%$ pristine (Pop III) stars and $50\%$ polluted stars, by mass. As $P_{\star}$ is known for each star particle, when a star enriches the surrounding medium through SNe, we can determine the fraction of Pop III SN ejecta contributed, allowing us to track the primordial metallicity contributed by these stars.  As such we are able to track the pollution of surrounding cells not only in terms of $\overline Z $ and $P$, but also in terms of the cells' primordial metals, $\overline Z_{\rm P}$. As discussed in detail below, this allows us to connect assumptions about the yields from Pop III stars with observations of carbon-enhanced metal-poor (CEMP-no) stars  \citep{2005ARA&A..43..531B, 2016A&A...586A.160H, 2016arXiv160706336Y},  which are defined as having [C/Fe] $>$ +0.7 and [Ba/Fe] $<$ 0.0 \citep{2005ARA&A..43..531B} . While \cite{2014ApJ...791..116C} and \cite{2016MNRAS.456.1410S} attribute the abundances in CEMP stars to primordial, faint SNe and anisotropic ejecta, respectively, our method depends only on mixing and the abundance patterns of Pop III SNe. 

By knowing the average metallicity, $\overline Z$, and the pristine gas fraction, $P$, we can better model the metallicity of the polluted fraction of gas (or stars). Since $\overline Z$ represents the average metallicity of a parcel of gas and the polluted fraction, $f_{\rm pol} \equiv 1-P$, models the fraction of gas that is actually polluted with metals, we can use the value of $f_{\rm pol}$ to predict the enhanced metallicity ($Z>\overline Z$) of the polluted fraction of gas in each simulation cell. This allows us to more accurately model the metallicity of the star particles created from this gas. 

However, our new scalar only indicates that the pristine fraction of the cell contains gas with $Z < Z_{\rm crit}$. When considering primordial cells we know that the pristine gas has $Z \approx 0$ and hence the polluted fraction, $f_{\rm pol}$, accounts for all of the metals in the cell. In this case we can use $f_{\rm pol}$ to precisely determine the enhanced metallicity of the polluted fraction of gas:

\begin{equation}\label{eq:zgas}
\begin{aligned}
Z \equiv \frac{\overline Z} {f_{\rm pol}},
\end{aligned}
\end{equation}
where $Z$ is the enhanced metallicity of the polluted fraction of gas.

For instance, consider an initially primordial parcel of gas that is enriched, on average, to $\overline Z = 10^{-1}\, Z_{\odot}$ with $P = 0.9 \implies f_{\rm pol}=0.1$ by a nearby SN event. The fraction of stars subsequently produced from such gas will, ignoring possible differences in the star-forming efficiency  between metal-free and polluted gas, be 90\% pristine and 10\% polluted. However, the metallicity of the polluted stars will be $10 \times$ greater than the average metallicity since all of the metals in the cell are currently concentrated in 10\% of its volume. Thus, in this case, we can simply correct the metallicity of star particles born in the gas by taking into account the polluted fraction of gas as described by Eq. \eqref{eq:zgas}.

However, once mixing has reduced $P < 1.0$ with $\overline Z > Z_{\rm crit}$ it becomes possible for the average metallicity of a cell to fall to $\overline Z < Z_{\rm crit}$ either via the advection of either pristine material from neighboring cells or via the movement of material with a lower average metallicity into the cell. Such cells' pristine fractions evolve back toward unity since mixing now dilutes areas with $Z > Z_{\rm crit}$ until the entire cell has $P = 1.0$ and $\overline Z = Z < Z_{\rm crit}$ throughout the cell. Now subsequent injections of pollutants can once again raise $\overline Z > Z_{\rm crit}$, but in this case not all of the metals are concentrated in the incoming pollutants.

\begin{figure}[h]
\begin{center}
\includegraphics[width=1\columnwidth]{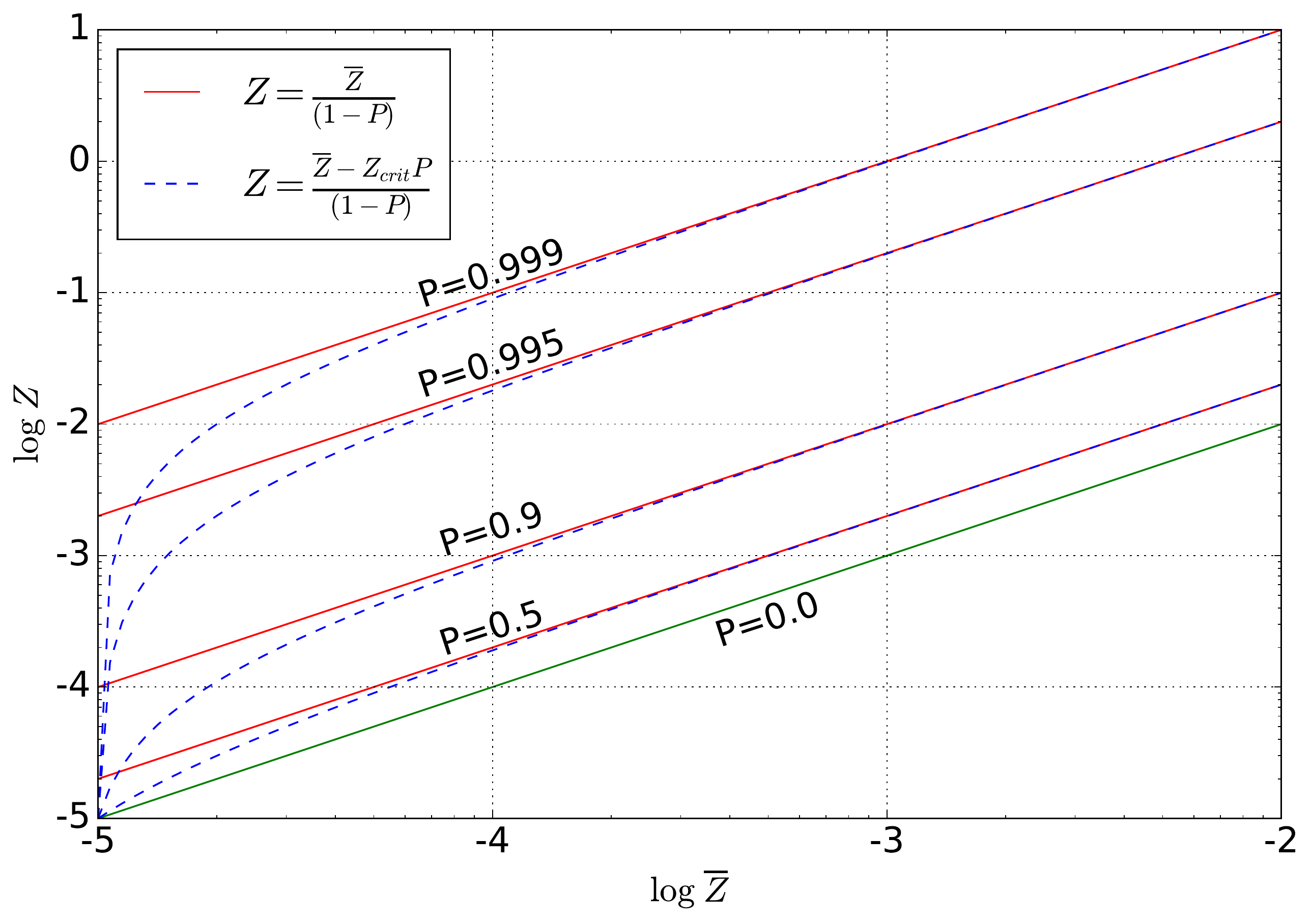}
\caption{Curves depicting the lower and upper bounds for the corrected metallicity, $Z$, of the polluted region of gas over a range of average cell metallicities, $\overline Z$. The corrected metallicity will fall on the solid line when the pristine fraction is primordial. The largest difference between the two corrections occurs at small polluted fractions, where pollutants are concentrated in a small volume, at low $\overline Z$. }
\label{fig:fpolBnd}
\end{center}
\end{figure}

Our scalar $P$ does not encode the metallicity of the pristine gas: the scalar only captures the fraction of gas with $Z < Z_{\rm crit}$ and as such we do not know the fraction of metals, captured in the scalar $\overline Z$, that are distributed in the pristine fraction of gas. However, we can bound this value since we know that the metallicity of the polluted and pristine fractions of the gas must sum to the average metallicity of the cell. Equation \eqref{eq:zquation} solves for the metallicity of the polluted fraction of gas, $Z$, when considering that the pristine fraction has an unknown metallicity $Z_{\Bbb P}$:

\begin{equation}\label{eq:zquation}
\begin{aligned}
\overline Z =&\, Z (1-P) + Z_{\Bbb P} P,\\
Z =&\, \frac{\overline Z - Z_{\Bbb P} P }{(1-P)}.
\end{aligned}
\end{equation}
It can be seen that when $Z_{\Bbb P} = 0$ we recover $Z = \overline Z/f_{\rm pol}$, the enhanced metallicity of a polluted volume of gas within a primordial cell. This correction is the upper bound for the metallicity of the polluted fraction. We can also establish a lower bound for the polluted fraction's metallicity correction by considering the definition of the pristine fraction, namely, that $Z_{\Bbb P} < Z_{\rm crit}$. Substituting for $Z_{\Bbb P}$ this means that

\begin{equation}\label{eq:lowerlim}
\begin{aligned}
Z = \frac{\overline Z - Z_{\rm crit} P }{(1-P)},
\end{aligned}
\end{equation}
is a lower bound on the correction to the metallicity of the polluted fraction. Comparing the lower and upper bounds of the correction to the metallicity of the polluted fraction, we see that the term $Z_{\rm crit} P$ will have the largest effect when $\overline Z \approx Z_{\rm crit}$ and $P \approx 1$. This is depicted in Figure~\ref{fig:fpolBnd}. 

The pristine gas fraction, $P$, is initialized to 1.0 for all cells in the simulation. As the first stars are formed and go SNe, metal-rich ejecta, along with entrained gas from the cell, is immediately carried into neighboring cells, where it increases the cell density from $\rho_{\rm cell}$ to $\rho_{\rm cell} + \rho_{\rm ejecta}$. This decreases the cell's pristine gas fraction from 1.0 to $P = \rho_{\rm cell}/(\rho_{\rm cell} + \rho_{\rm ejecta})$ starting the decay of $P$. We note that this method of computing the change to $P$ assumes that the ejecta (with density $\rho_{\rm ejecta}$) is well mixed. The subgrid mixing algorithm is then invoked for each cell at each time-step, resulting in the decay of the pristine gas fraction (based on theoretical modeling of the pristine fraction with physical parameters calibrated by numerical simulations, as described below) for all cells where $P < 1.0$.

Finally, we note that star formation is likely to be more efficient in polluted gas, due to more efficient cooling. However, given the uncertainties in the Pop III initial mass function (IMF) we have elected to weigh star formation rates in pristine and polluted gas only by the pristine gas fraction. For example, stars born in gas with $P = 0.9$ give rise to star particles with $P_{\star} = 0.9$ meaning 90\% of the mass of the star particle represents Pop III stars. Below we include Table~\ref{table:notation} clarifying the notation used to describe the scalars we reference throughout this paper.

\begin{deluxetable}{cp{0.65\linewidth}}
\tabletypesize{\footnotesize}
\tablecolumns{2} 
\tablecaption{Notation} 
\tablehead{\colhead{Symbol} & \colhead{Definition} } 
\startdata
\multirow{2}{*}{$P$, $P_{\star}$} & The mass fraction of gas (star particles) with $Z < Z_{\rm crit}$. \\ \hline \\ [-1.5ex]
\multirow{2}{*}{$f_{\rm pol}$} & The mass fraction of gas or star particles with $Z \ge Z_{\rm crit}$. $f_{\rm pol} \equiv 1-P$.  \\ \hline  \\ [-1.5ex]
\multirow{3}{*}{$Z_{\Bbb P}$} & The metallicity of the pristine fraction of gas or star particle; while undetermined, it is bounded by $0 \le Z_{\Bbb P} \le Z_{\rm crit}$.  \\ \hline  \\ [-1.5ex]
\multirow{3}{*}{$\overline Z$, $\overline Z_{\rm P}$, $\overline Z_{\star}, \overline Z_{\rm P, \star}$} & The average metallicity (primordial metallicity) within an unmixed volume of gas (or star particle). \\ \hline  \\ [-1.5ex]
\multirow{4}{*}{$Z$, $Z_{\rm P}$ $Z_{\star}, Z_{\rm P, \star}$} &  The corrected and assumed homogeneous metallicity (primordial metallicity) of the polluted fraction of gas (or star particle).
\enddata \label{table:notation}
\tablenotetext{}{} 
\end{deluxetable} 

\subsection{Subgrid Model for the Pollution of Pristine Gas}

\subsubsection{Self-Convolution Model}

The novelty of our approach is in how the pollution of pristine gas is modeled on subgrid scales. Here we rely on the work described in PS13, which determined that in fully developed turbulence, the pollution rate can be modeled by a relatively simple self-convolution model. While SN forcing is intuitively compressive, P13 used solenoidal forcing to drive turbulence. However, studies by \cite{2010A&A...512A..81F} and \cite{2016ApJ...822...11P} demonstrate that SN driving forces are not purely compressive. Indeed, the effective SN driving force may possibly be more solenoidal than compressive with a compressive-to-solenoidal ratio below one, due to the non-sphericity of the SN blast which tends to be clumpy in both velocity and density space.  Additionally, each `SN' in our simulation represents the combined energy of several SN since each star particle represents a Salpeter IMF with a total mass in the range $726-3628\, M_{\odot}$. These considerations conspire to make the ISM highly turbulent. While compressive forcing is surely a part of SN-driven turbulence, results from simulations using solenoidally driven turbulence are a reasonable approximation when determining the mixing timescales for SN-driven turbulence. 

There has been compelling evidence that in a turbulent medium, the dominant scalar structures at small scales are two-dimensional sheets or edges \citep[e.g.][]{2011PhRvE..83d5302P}, and the rate at which these sheets are produced is determined mainly by the turbulent stretching rate at large length scales. With time, the sheets become thinner, and once their thickness is sufficiently small for molecular diffusivity to efficiently operate, neighboring sheets are homogenized, leading to a reduction in the width of the local metallicity probability distribution function, $\Phi(Z;t)$.

The evolution of $\Phi(Z;t)$ within a turbulent region in this physical picture can be approximately described as
\begin{equation} \label{curl}
\begin{aligned}
\frac{\partial \Phi(Z; t)}{\partial t} = s(t) \left[ - \Phi(Z; t) + \int\limits_{0}^{1} dZ_{1} \Phi(Z_{1};t) \times \right. \\ 
\left. \int\limits_{0}^{1} dZ_{2} \Phi(Z_{2}; t)  \times  \delta \left( Z-\frac{Z_{1} +Z_{2}}{2} \right) \right], 
\end{aligned}
\end{equation}
where $s(t)$ is the turbulent stretching rate that controls the rate at which the probability density function (PDF) convolution proceeds \citep{1963Curl,1979PhFl...22...20D,1979JNET....4...47J}.  Extending this model, \cite{2007PhFl...19b8101V} developed a ``continuous'' version, which essentially assumes that the convolution occurs everywhere in the flow at any given time, but in an infinitesimal time, $\Delta t$, the number of convolutions is infinitesimal and equal to $\epsilon = s(t) \Delta t$.  \cite{2008JFM...617...51D} then generalized this even further, including a parameter $n$ such that a fraction $n \epsilon$ of the flow experiences mixing events during a time interval, $\Delta t$, and the number of convolutions in this fraction 
of the flow is 1/$n.$ In this model, $n$ characterizes the degree of spatial locality of the PDF convolution, with larger values of $n$ corresponding to more global convolutions. The models of \cite{1963Curl} and \cite{2007PhFl...19b8101V} correspond to $n=1$ and $n\to \infty$, respectively.

By integrating $\Phi(Z;t)$ from a finite but extremely small value up to the critical metallicity, the \cite{2008JFM...617...51D} model can be used to derive a very simple equation for the evolution of the $Z<Z_{\rm crit}$ pristine fraction:  
\begin{align}\label{eq:selfConv}
\frac{d P}{d t} = - \frac{n}{\tau_{\rm con}}  P(1-P ^{1/n}). 
\end{align}
The $n=1$ case of this equation was first given in \cite{2007ApJ...654L..29P}. This equation traces the evolution of $P$ as a function of $n$ and a timescale $\tau_{\rm con}$, and these parameters, in turn, are functions of the turbulent Mach number, $M$, and the average metallicity of the cell relative to the critical metallicity, $\overline Z /Z_{\rm crit}$ \citep[][PS13]{2012JFM...700..459P}. 

Note that Eq. \eqref{eq:selfConv} has the property that if $P = 1$ then $\dot P = 0$ ensuring that pristine cells remain pristine.  There are no metals in such a cell with which to pollute it.  As soon as  the injection of polluted material causes $P < 1.0$ however, the polluted fraction will then continue to decrease. 
Including the cell-to-cell advection handled by \textsc{ramses} and the addition of enriched ejecta material to a cell, the full equation for the evolution of $P$ is 
\begin{align}\label{eq:mix1}
\frac{\partial (\rho P)}{\partial t}  + {\bf \nabla} \cdot ({\bf u} \, \rho \, P)
= -\frac{n}{\tau_{\rm con}}  \rho P (1 - P^{1/n}) - \dot \rho_{\rm ej} P,
\end{align}
where $\rho$ and ${\bf u}$ are the local density and velocity, and  $ \dot \rho_{\rm ej}$ is the rate that the density of cell is increased by the addition of ejecta.  The reader may notice that we have omitted the diffusion term described by PS13 (Eq. (49) in that work). We have not tried to characterize the numerical diffusion inherent in \textsc{ramses} and a proper treatment would need to account for any difference between it and the diffusion term in PS13 when computing cell-to-cell diffusion. We leave this for a future work.

\subsubsection{Locality Parameter and Convolution Timescale}\label{sec:local}

The evolution of $P$ described in Eqs.\ \eqref{eq:selfConv} and \eqref{eq:mix1} depends on a convolution timescale, $\tau_{\rm con}$ and the parameter \textit{n}, which quantifies the locality of mixing.
Here we are interested in the case in which the driving scale of the turbulence and the length scale at which pollutants are added to the medium (referred to as $L_f$ and $L_p$ in PS13) both occur on the grid scale
$\Delta x.$   In this case, as shown in PS13, the locality parameter as a function of Mach number is well fit by
\begin{equation}\label{eq:n}
n  = \, 1+11\, \exp \left(-\frac{M}{3.5} \right).
\end{equation}
This means that in subsonic turbulence, pollution is more of a global process, corresponding to $n \approx 12,$ and in highly supersonic turbulence pollution is more local, corresponding to $n \approx 1$,  the $n$ value in Curl's original model.

Also following PS13, we model the dynamical mixing time as 
\begin{equation}
\tau_{\rm con} = \frac{\Delta x}{v_t} 
\begin{cases} \tilde \tau_{\rm con1}&\mbox{if } P \ge 0.9 \\ 
\tilde \tau_{\rm con2}& \mbox{if }  P < 0.9 \end{cases},
\end{equation}
where $v_t$ is the turbulent velocity of the cell, and $\tilde \tau_{\rm con1}$ and $\tilde \tau_{\rm con2}$ account for the slightly less efficient mixing that occurs at low pollution fractions (PS13).
Unlike $n$, these timescales depend not only on the Mach number but also on the ratio of the average metallicity in the turbulent region to the critical metallicity. For example, if the mean metallicity in the region is less than the critical metallicity, then the fraction of pristine material cannot monotonically decrease since complete mixing with $\overline Z < Z_{\rm crit}$ should imply $P = 1.0$. 
On the other hand, as PS13 were concerned only with cases in which the mean metallicity in the medium was much greater than the critical metallicity, their fits to the convolution timescale show only a weak dependence on $\overline Z/Z_{\rm crit}$.  Using an extrapolation of the dependence $\overline Z/Z_{\rm crit}$ given in PS13 for $\overline Z/Z_{\rm crit} < 1.0$ results in $P$ always decreasing -- even when the average metallicity is subcritical.

\begin{figure}[h]
\begin{center}
\includegraphics[width=1\columnwidth]{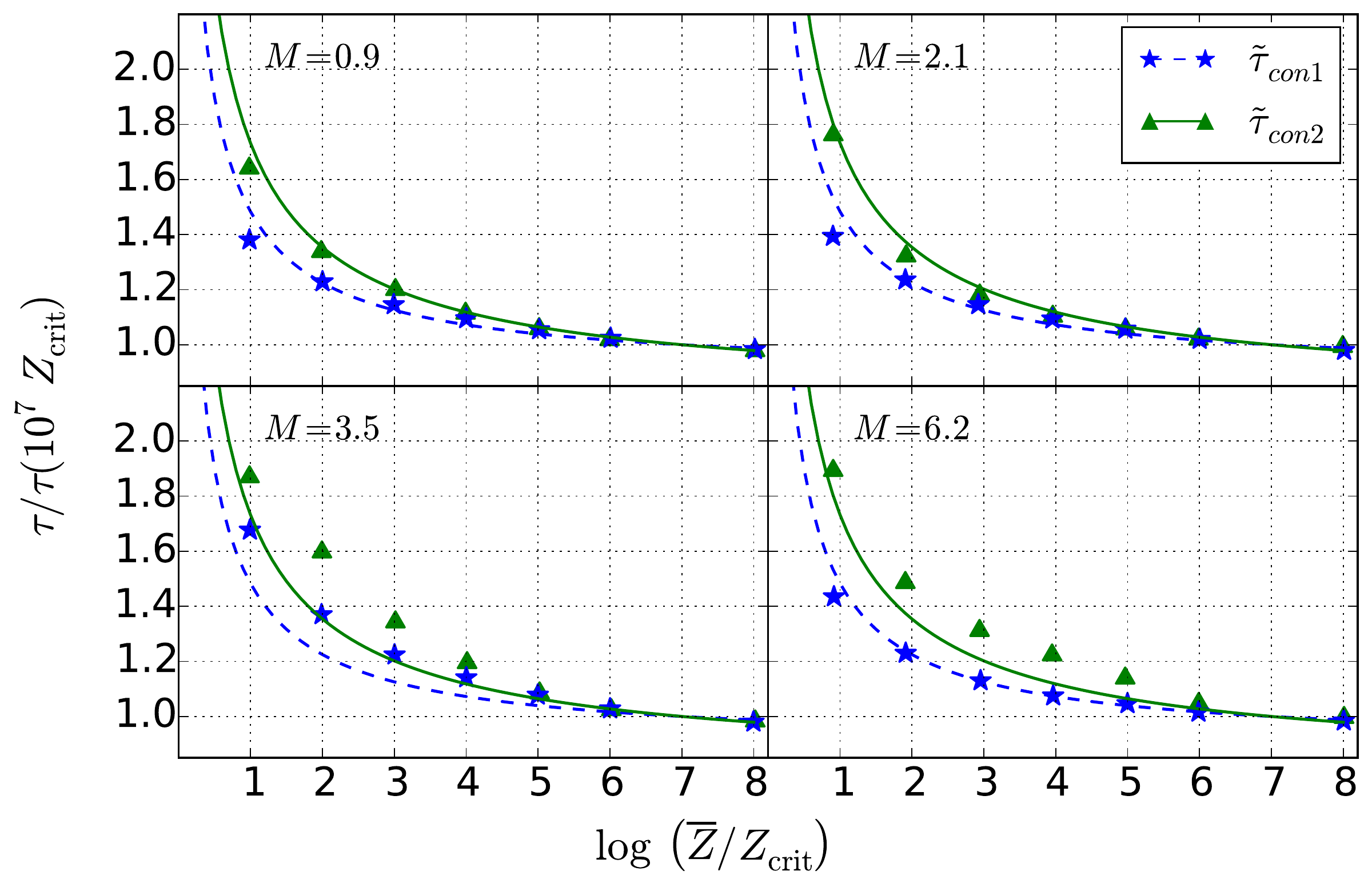}
\caption{Our analytic fits to the dynamical mixing times ($triangles$ and $stars$) as determined by PS13. Time scale $\tilde \tau_{\rm{con1}}$ is used when $P \ge 0.9$, $\tilde \tau_{\rm{con2}}$ is used otherwise. For this plot, the mixing time is normalized to the mixing time for region with $Z_{\rm crit} = 10^{-7}\, \overline Z$.  Note that as $\overline Z  \to Z_{\rm crit}$ the mixing time becomes exponentially longer, going to infinity when $\overline Z < Z_{\rm crit}$.}
\label{fig:tau}
\end{center}
\end{figure}

Thus, we refit the dependence of $\tilde\tau_{\rm con1},$ $\tilde\tau_{\rm con2}$ on $\overline Z/{Z_{\rm crit}}$ working with the results of the PS13 data over a much larger range of values.
Figure~\ref{fig:tau} depicts the normalized convolution timescale as a function of $\overline Z $ for the range of turbulent flows studied by PS13. For convenience we normalize to  its value when $Z_{\rm crit} = 10^{-7}\, \overline Z$ as this value was used in the fits presented in PS13.  As expected, the mixing time increases exponentially as $\overline Z $ approaches $Z_{\rm crit}$.  
As the critical metallicity in the simulations in PS13 was taken to be $10^{-7}$ of the mean metallicity, we define our fits in terms of the ratio
\begin{equation}
x\equiv-\log_{10} \left(\frac{\overline Z}{10^7 Z_{\rm crit}}\right) \left[{\log_{10} \left(\frac{\overline Z}{Z_{\rm crit}}\right)}\right]^{-1}, 
\end{equation}
which equals zero when $\overline Z/{Z_{\rm crit}}  = 10^7$ and approaches infinity as $\overline Z/Z_{\rm crit}$ drops to 1.  We then obtained a simple fit to the PS13 simulation results as
\begin{equation}\label{eq:tau}
\begin{aligned}
\tilde \tau_{\rm con1} = &\left[0.225 - 0.055\,\exp\left(-\frac{M^{3/2}}{4} \right) \right] \, \sqrt{\frac{x}{5}+1}, \\
\tilde \tau_{\rm con2} = &\left[0.335 - 0.095\,\exp\left(-\frac{M^{2}}{4} \right) \right] \, \sqrt{\frac{x}{3}+1}, \\
\end{aligned}
\end{equation}
if $\overline Z/Z_{\rm crit} > 1$ and $\tau_{\rm con} = \infty$ if $\overline Z/{Z_{\rm crit}} \le 1.$  
It is possible for a region with $\overline Z > Z_{\rm crit}$ to evolve back to $\overline Z \le Z_{\rm crit}$ due to the advection of highly pristine material from nearby cells. In this case we think of mixing as making the cell more pristine, rather than less, and evolve the pristine gas fraction back toward 1 as an exponential function of the cell's turbulent velocity:
\begin{equation}\label{eq:Pback}
\frac{d}{dt} (1-P )=  - (1 - P) \frac{v_t}{\Delta x}.
\end{equation}

We note that the pristine fraction does not always evolve toward 1 immediately once $\overline Z  < Z_{\rm crit}$. For example, if  
the polluted fraction in a cell  is below $\overline Z/ Z_{\rm crit}$, the average metallicity in the polluted region 
is still above $Z_{\rm crit}$, and thus the pristine fraction in the cell would still be decreasing as the polluted region mixes with more ambient pristine gas.  
Roughly speaking, only after the polluted fraction in a cell exceeds $\overline Z/ Z_{\rm crit}$, does the pristine fraction start to evolve toward 1. 
However, accounting for this complexity by checking whether the polluted fraction in a cell is above or below $\overline Z/ Z_{\rm crit}$ 
does not cause a significant  difference from the results using Eq. \eqref{eq:Pback} immediately (when the average metallicity in a cell drops below  $Z_{\rm crit}$). 
We will therefore focus on results from equation \eqref{eq:Pback}.

All the fits above depend on the cells' turbulent velocity, $v_t$ and its ratio to the local sound speed, $M=v_t/c_s.$ 
This velocity can be estimated in turn as $v_{\rm t} = \nu_{\rm t}/\Delta x,$ where $\nu_{\rm t}$ is the turbulent kinematic eddy viscosity of the scale of the cell, or as $v_{\rm t} = \sqrt{2 K}$ where K is the subgrid kinetic energy. Many possible models exist in the literature for the estimate of $v_t$ \citep[e.g.][]{1986PhFl...29.2152Y,1991PhFl....3.2746M,1992JFM...238..155E,1997JFM...339..357V} 
or $K$ \citep[e.g.][]{1975JCoPh..18..376S,1984JAtS...41.2052M,1995JFM...286..229G,2006A&A...450..265S,2010JTurb..11....4G,2010MNRAS.405.1634S,2012JFM...699..385C}.  
While a comparison between different approaches merits further study, here we adopt the simplest approach, making use of the eddy viscosity model of \cite{1963MWRv...91...99S}. A brief overview of the approach used to compute $v_t$ follows.

We first compute the numerical velocity gradients across each cell $\Delta_{\rm i} v_{\rm j}$ to determine the local rate-of-strain tensor  
\begin{equation}\label{eqn:S}
\begin{aligned}
\sc{S}_{\rm ij}\equiv \frac{1}{2}(\Delta_{\rm i} v_{\rm j} + \Delta_{\rm j} v_{\rm i}),
\end{aligned}
\end{equation}
which captures the 3D velocity shear around each cell 
\citep{2014ApJ...784...94S}. Starting with the energy in the Kolmogorov inertial spectrum ($\epsilon^{2/3} k^{-5/3} $), and equating it to the loss of kinetic energy in the flow, $2 \nu \langle \sc{S}_{ij} \sc{S}_{ij}\rangle$, we have the following: 
\begin{equation}
2 \nu \langle \sc{S}_{ij}  \sc{S}_{ij}\rangle = 2 \epsilon^{2/3} \int_{0}^{1/\Delta x}{k^2\, k^{-5/3} dk} \propto \epsilon^{2/3} \Delta x^{-4/3},
\end{equation}
where we have summed the energy spectrum (in k-space) up to the size of the filter scale to capture the sub-grid energy. 
Noting that $\epsilon \propto v_t^3/\Delta x$, results in
\begin{equation}
v_t \propto  \sqrt{\nu}\;|\overline \sc{S}_{ij}|\; \Delta x,
\end{equation}
demonstrating that the amplitude of velocity fluctuations on the filter scale is directly related to the magnitude of the rate of strain, as adopted in our simulation and by Smagorinsky: $v_t = |\overline \sc{S}_{ij}|\; \Delta x$.

\subsection{Molecular Cooling}

\begin{figure}[h]
\begin{center}
\includegraphics[width=1\columnwidth]{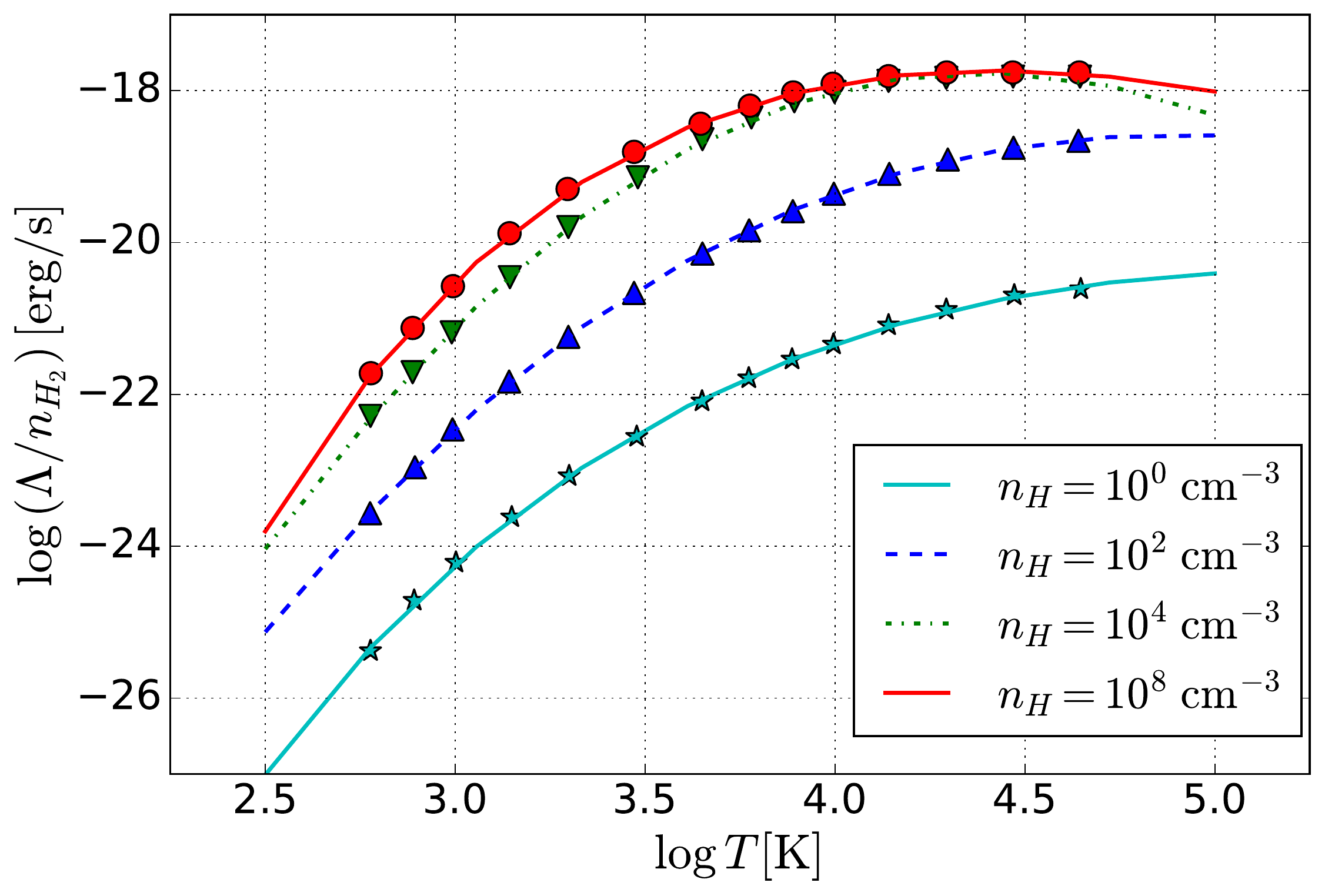}
\caption{Cooling per $H_{\rm 2}$ molecule. Data points are from \cite{1996ApJ...461..265M}. The curves are our analytic fits and are used in the code.}
\label{fig:coolingcurve}
\end{center}
\end{figure}

Beyond the subgrid model for the pollution of pristine gas described above, we have also modified \textsc{ramses} to include a simple molecular cooling model which is important for low-temperature cooling in the pristine gas \citep{,2006MNRAS.366..247J, 2008arXiv0809.2786P, 2013ApJ...763...52H}. Our model is analytic and based on the work of \cite{1996ApJ...461..265M}, which provides a radiative cooling rate per $H_{\rm 2}$ molecule, $\Lambda_{r}/n_{H_2}$, across a range of densities, as depicted in Fig.~\ref{fig:coolingcurve}.  We truncate our $H_{\rm 2}$  cooling model above 50,000 K as the contribution of molecular cooling becomes negligible and molecular hydrogen is highly dissociated above this temperature.

The cooling rate is computed for each simulation cell based on the cell's density, temperature, and molecular fraction, $f_{\rm{H_2}}$.
Our initial $H_{\rm 2}$ fraction is primordial \citep[$f_{\rm{H_2}}=10^{-6}$;][]{2005MNRAS.363..393R}, we model the Lyman-Werner flux from our star particles as $\eta_{\rm{LW}} = 10^{4}$ photons per stellar baryon \citep{2006MNRAS.373..128G} and we assume optically thin gas throughout the simulation volume. We compute the number of stellar baryons, $N_{*,b}$, by totaling the mass in star particles, at each simulation step, assuming a near-primordial composition ($X$=0.73, $Y$=0.25). This results in an updated $f_{\rm{H_2}}$ for each simulation step:

\begin{equation}\label{eqn:fh21}
f_{H_2,new} = \frac{(f_{H_2,old} \; N_{gas} - N_{\rm LW})}{N_{gas}},
\end{equation}
where\\ 
\begin{equation}\label{eqn:fh22}
N_{\rm LW} = N_{*,b} \; \eta_{\rm LW}.
\end{equation}

We note that the total stellar mass created at the onset of star formation generates enough Lyman-Werner flux to permeate our simulation volume of 27 Mpc$^{3}$ h$^{-3}$ \citep{2013MNRAS.428.1857J, 2003Natur.425..812B}, destroying all of the molecular hydrogen. We do not model subsequent $H_{\rm 2}$ formation since cooling becomes dominated by metal lines shortly after the first star particles form and any subsequent molecular hydrogen would be quickly destroyed by the Lyman-Werner flux. Hence, our molecular cooling model is significant only for the very first generation of stars.  

Metal-line cooling is computed as a function of gas metallicity and temperature using \cite{1993ApJS...88..253S} for temperatures down to $\approx 10^4\, K$ below which \textsc{ramses} makes use of \cite{1995ApJ...440..634R}. We fix the gas temperature floor at 100 K for radiative cooling although adiabatic cooling below this limit is possible. Lastly, our UV background is based on the work of \cite{1996ApJ...461...20H}. 

\subsection{Star Formation and SN Generation}\label{sec:sfsn}

Following the cosmic evolution of pristine gas is sensitive to our implementation of star formation and SN generation.
\textsc{ramses} creates star particles in regions of gas according to a Schmidt law \citep{1959ApJ...129..243S},  as
\begin{equation}\label{eqn:sf}
\frac{d\rho_{\star}}{dt} = \frac{\rho}{t_{\star}} \theta(\rho- \rho_{\rm th}),
\end{equation}
where the Heaviside step function $\theta(\rho- \rho_{\rm th})$ allows for star formation only when the density exceeds a threshold value $\rho_{\rm th}.$  Here we have set $\rho_{\rm th}$ to be the maximum of $0.75 \, m_p \, {\rm cm}^{-3}$ and 150 times the mean density in the simulation, where the latter criterion ensures that star particles are only formed in virialized halos and not in high-density regions of the cosmological flow \citep{2006A&A...445....1R, 2008A&A...477...79D}.  To generate an SFR in $M_{\odot} \, {\rm yr}^{-1}$ in good agreement with \cite{2014ARA&A..52..415M} and \cite{2012ApJ...754...83B}, we have tuned the star formation time scale $t_{\star} = 0.316$ Gyr, approximately 5 times the free fall time, $t_{\rm ff} \equiv \left(3 \pi /32 G \rho \right)^{1/2},$ for a gas with $\rho = 0.75\, m_p \,  {\rm cm}^{-3}.$
The mass of the newly created star particle is $m_{\star} = \rho_{\rm th}\Delta x_{\rm min}^3 N$, where $\Delta x_{\rm min}$ is the best resolution cell size and N is drawn from a Poisson distribution
\begin{equation}\label{eqn:poi}
P(N) = \frac{\overline N }{N!}\, \exp(-\overline N ),
\end{equation}
with
\begin{equation}\label{eqn:poiN}
\overline N  = \frac{\rho\, \Delta x^3}{\rho_{\rm th}\Delta x_{\rm min}^3 } \frac{\Delta t}{t_{\star}}.
\end{equation}
A further limitation on star particle formulation is that no more than 90\% of the cell's gas mass can be converted into stars.

As each star particle represents a mass range of stars, a fraction of the mass will be returned to the grid in the form of SN.   In our simulations, this recycling is assumed to occur after  the 10 Myr lifetimes for stars near the top of the IMF \citep[e.g.][]{2008ApJ...689..358R}.  In this case the impact of massive stars is determined by the fraction of the star particle mass they eject, $\eta_{\rm SN},$ and the kinetic energy per unit mass of this ejecta, $E_{\rm SN}$.  For simplicity, we take $\eta_{\rm SN} = 0.1$ and $E_{\rm SN} = 1 \times 10^{51}\, {\rm ergs}/10\, M_{\odot}$, for all stars formed throughout the simulation, regardless of their primordial fractions.   Note however that as the Pop III IMF is likely to have been biased to massive stars, metal ejection from such stars may have been more efficient \citep[e.g.][]{2003ApJ...589...35S,2005ApJ...624L...1S}, leading to differences in stellar enrichment which we plan to explore in future work.

For each newly formed star particle, the ejected mass and energy are deposited into all cells whose centers are within 150 pc and if the size of the cell containing the particle is greater than 150 pc, the energy and ejecta are deposed into the  adjacent cells \citep{2008A&A...477...79D}. Here the total mass of the ejecta is that of the stellar material plus an amount of the gas within the cell hosting the star particle (\textit{entrained gas}) such that 
$m_{\rm ej} =  m_{\rm sn} + m_{\rm ent}$, with $m_{\rm sn} \equiv \eta_{\rm sn}\, m_{\star}$ and $m_{\rm ent} \equiv {\rm min}(10\, m_{\rm sn},\; 0.25\, \rho_{\rm cell}\, \Delta x^{3})$.  Similarly, the mass in metals added to the simulation is taken to be 15\% of the SN ejecta plus the metals in the entrained material, $Z_{\rm ej}\; m_{\rm ej} = m_{\rm ent}\,Z + 0.15\, m_{\rm sn},$ and the mass in primordial metals is taken to be  $Z_{\rm P, ej}\; m_{\rm ej} = m_{\rm ent}\,Z_{\rm P} + 0.15\, m_{\rm sn}\, P_{\star}.$  SN energy is the dominate driver of turbulence in our simulation and we have chosen to partition it equally between kinetic and thermal energy. Lastly, we note that we do not model black hole formation or feedback.

Since our SN feedback model deposits SN energy in the surrounding cells, it tends to leave the central portion of the star-forming cloud, in the host cell, mostly intact. Additionally, while radiative feedback, especially from massive stars, can be quite effective in evacuating gas and disrupting star formation in small halos \citep{2012AIPC.1480..123W, 2004ApJ...610...14W}, we have not modeled it in this simulation. 
Conversely, however, the ionization and shock fronts created by radiative feedback may also trigger star formation in clumps or pillars \citep{2012A&A...546A..33T, 2010A&A...523A...6D} of very dense gas that are likely to form near the shock boundary. Given these considerations we feel our approach is a reasonable starting point for modeling star formation in turbulent flows. We leave the analysis of the impact of radiative feedback and different SN energies for future study. 

\subsection{Simulation Setup}

We evolved a cubic 3 Mpc $h^{-1}$ comoving box from $z=499$ to $z=5$ starting from initial conditions generated by the
\textsc{mpgrafic} code  \citep{2008ApJS..178..179P}.  The initial gas metallicity was $Z=Z_{\rm P} = 0,$ the initial $H_{\rm 2}$ fraction was $10^{-6},$ and we define $Z_{\rm crit} = 10^{-5} Z_\odot.$  The base resolution was $512^{3}$ cells ($\textit{l}_{\rm min} = 9$) corresponding to a  grid resolution of 5.86 comoving kpc h$^{-1},$ and a dark matter particle mass of $5.58 \times 10^{4}\, h^{-1}\, \Omega_{\rm \textsc{dm}}\, M_{\odot}.$   We refined cells as they become $8\times$ overdense, or when the local Jeans length is less than four times the current cell size, ensuring we always resolved the Jeans length with at least 4 simulation cells. We allowed for up 8 additional refinement levels ($\textit{l}_{\rm max} = 17$), using a courant factor of 0.8, resulting in a best possible spatial resolution of 22 physical pc $h^{-1}$. However, because these additional levels maintain a maximum physical resolution rather than a comoving resolution, the highest refinement level reached by $z=5$ was level 14.  Our settings resulted in a range of star particle masses between $726\, M_{\odot} \leq M \leq 3628\, M_{\odot}$.  The nonlinear length scale at the end of the simulation was $0.15$ $h^{-1}$ comoving Mpc, corresponding to a mass of $1.5 \times 10^9  M_\odot.$ Finally, we tuned the code reionization parameters to ensure that the reionization redshift agrees with recent results \citep[specifically, $z_{\rm reion} \approx 8.8$;][]{2015arXiv150201589P}.

\section{Results}

\subsection{Evolution of the Overall Star Formation Rate}

\begin{figure}[t]
\begin{center}
\includegraphics[width=1\columnwidth]{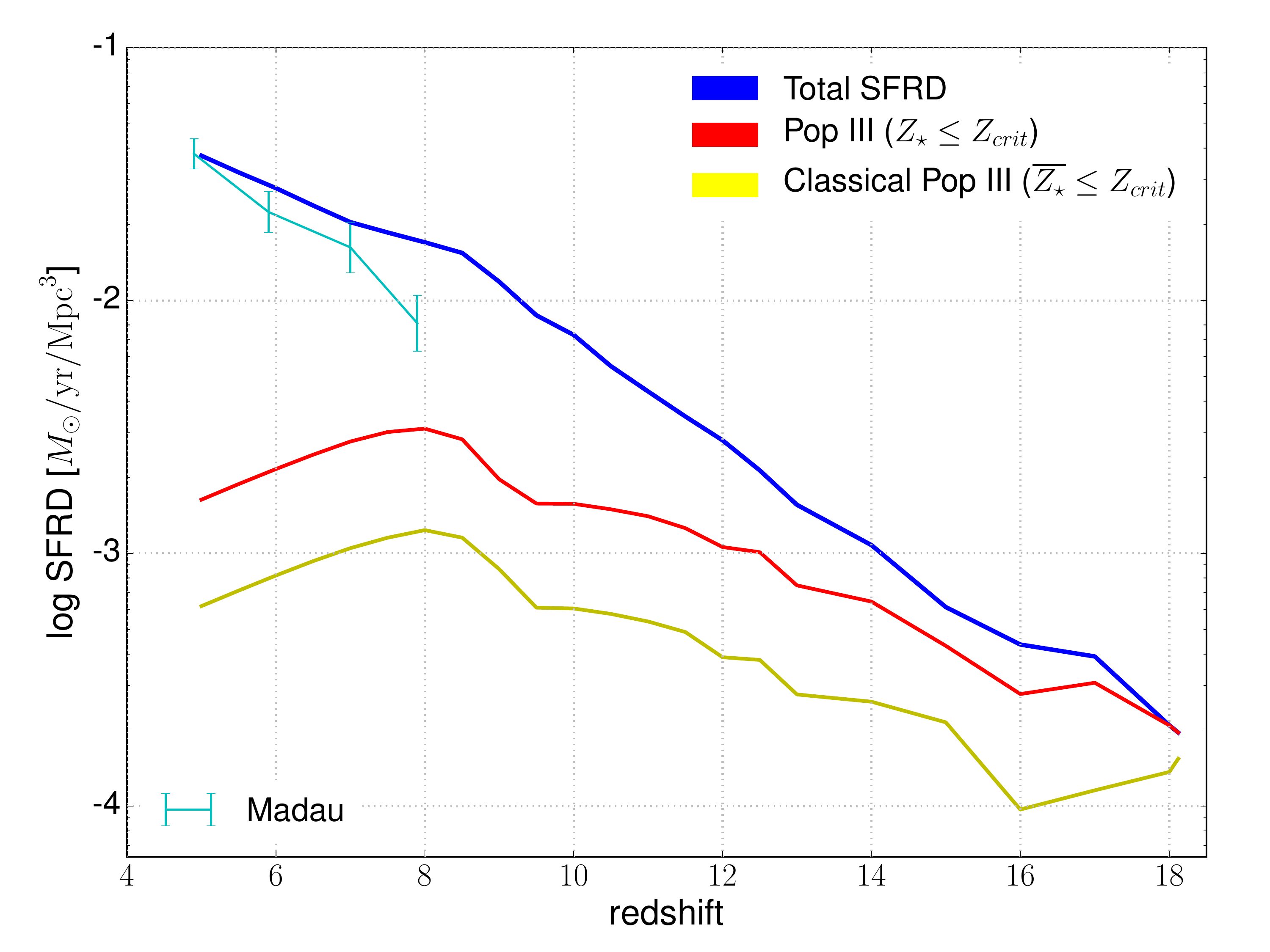}
\caption{Star formation rate (SFR) density for our simulation. For this plot, the ``classical Pop III" rate includes only those stars formed in cells in which  $\overline Z < Z_{\rm crit}$. The Pop III curve also includes metal-free stars found in star particles with $\overline Z >Z_{\rm crit}$: the star particle mass fraction $P_{\star} \times m_{\star}$. Note that the overall Pop III SFR density is $2-3\times$ higher than the classical rate, indicating that roughly $2/3$ of Pop III stars form in partially polluted regions with a mean metallicity $\overline Z >Z_{\rm crit}$. We also show the overall observational SFR density compiled by \cite{2014ARA&A..52..415M} from \citep{2012ApJ...752L...5B,2012ApJ...754...83B}. }
\label{fig:sfr}
\end{center}
\end{figure}

Our simulation  generated a total of 345 halos (at z=5) with a mass range of 79,000 to 372,468,025 $M_{\odot}$. The largest halo was composed of 513,749 star particles, while the smallest considered had 109. Halos needed to consist of at least 100 star particles before we counted them in this total. Our star formation over-density threshold ($150\times$ overdense), ensured that star formation only occurred in and around collapsed objects.  The mean matter density of our simulation in terms of the average matter density of the universe was $\langle \rho_{\rm sim} \rangle/(\rho_{\rm crit} \Omega_{\rm M})= 1.001$ indicating our volume represents a typical region of the universe.

Figure~\ref{fig:sfr} shows the evolution of the star formation rate (SFR) density in our simulation, along with observational results compiled by \cite{2014ARA&A..52..415M} from \cite{2012ApJ...752L...5B, 2012ApJ...754...83B}. The onset of star formation occurs at $z \approx 18,$ which quickly dissociates the small initial level of $H_2$ in the box, raising the minimum virial temperature of halos in which stars can form efficiently to $\approx 10^4$K.  This is above the nonlinear scale for redshifts above $z \approx 6,$ and so the SFR density increases roughly exponentially with decreasing redshift throughout the simulation, with some flattening below the redshift of reionization \citep{2005ApJ...624L...1S}.  This overall rate is also in good agreement with  $z \leq 7$ observations, and while our results are somewhat higher than observations at $z=8,$ this is to be expected as galaxy surveys are magnitude limited, while our simulations also include star formation down to very small mass limits.
 
For comparison, our high-redshift SFR density is much lower than the one obtained from the Renaissance zoom Simulation of a 'normal' region of the universe \citep{2016arXiv160407842X}. However, their simulation retains a maximum resolution of 19 comoving pc and ours maintains a maximum resolution of 23 physical pc. At their final simulation redshift  of $z \approx 12.5$, this corresponds to a physical resolution of $\approx$ 1.5 pc vs our $\approx$ 23 pc, allowing them to trace star formation in much smaller objects than we track and discuss here. 

While the very first stars in our simulation are purely metal-free (Pop III),  they quickly generate SNe, and the resulting pollution of the pristine gas gives rise to subsequent stellar populations with varying levels of metals. As our subgrid models allows us to track the pristine gas fraction for every particle, we can calculate not only the ``classically determined" Pop III star formation rate in $\overline Z < Z_{\rm crit}$ regions but also the formation of stars in unmixed regions with $\overline Z > Z_{\rm crit}$ as $P_{\star} \times m_{\star}$. In the classically determined case, our results are in good agreement with the simulations of \cite{2014MNRAS.440.2498P} \citep[see also][]{2007MNRAS.382..945T} over the range $5 \leq z \leq 8,$ while at higher redshifts our rate is somewhat higher, most likely due to our higher base resolution (5.86  versus 19.53 comoving kpc $h^{-1}$). This extended evolution is indicative of a large-scale spatially-inhomogeneous Pop III/Pop II transition as progressively lower-sigma peaks collapse and form stars in regions far away from sites of previous metal enrichment \citep{2003ApJ...589...35S}.

On the other hand, including metal-free star formation in unmixed regions with $\overline Z > Z_{\rm crit}$ leads to an increase in the Pop III rate by a factor of 2-3 over the full range from $5 \leq z \leq 16.$ This indicates that small-scale mixing is at least as important as large-scale inhomogeneities in determining the history of metal-free star formation. Or, in other words, at least as many Pop III stars are formed within metal-enriched protogalaxies as are formed in purely metal-free objects.

As noted earlier, our SFR density is dependent on the stellar and SN feedback prescriptions we have adopted, which are the standard ones in \textsc{ramses}. In particular, we have not modeled radiative feedback from these first, massive stars. As discussed in \cite{2004ApJ...610...14W}, such feedback is likely quite effective in dispersing the original star-forming cloud. We leave the modeling of radiative feedback -- as well as progenitor-dependent SN energy -- to a follow-up work.  

\subsection{The Gas}\label{sec:thegas}

\begin{figure*}
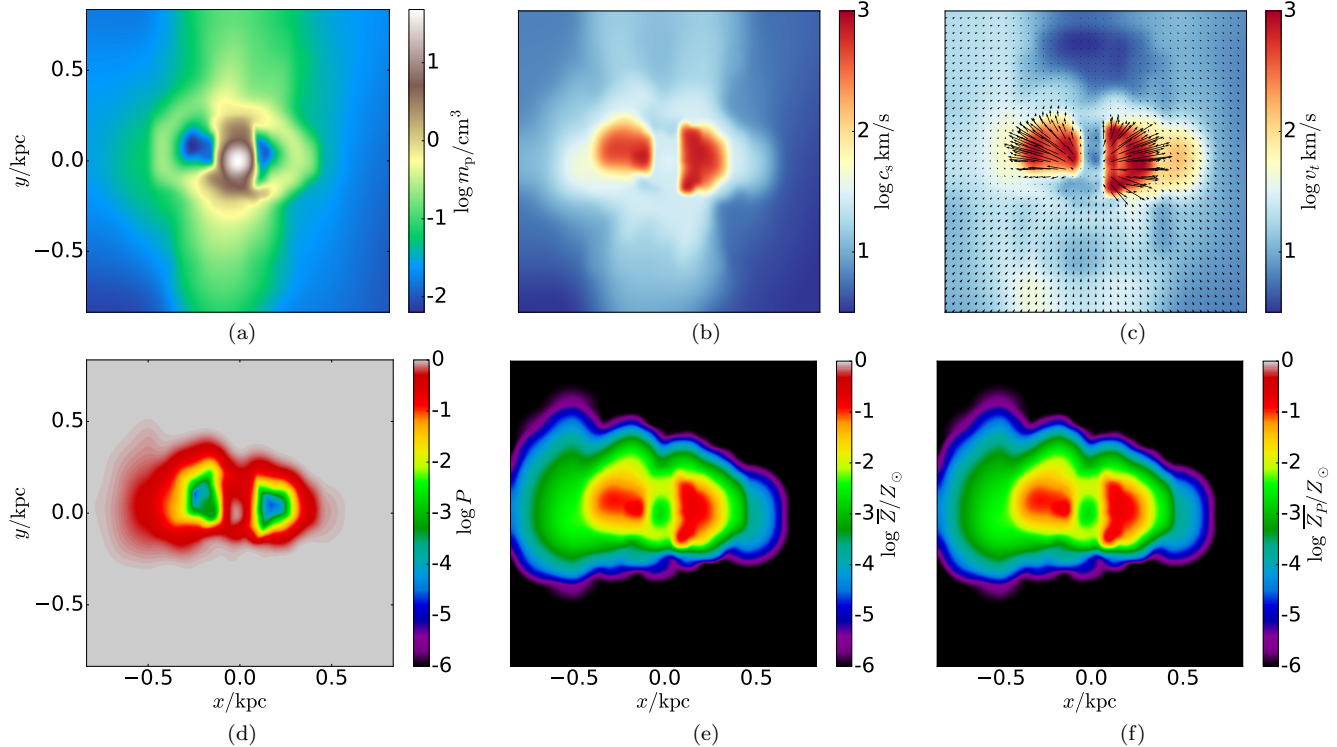

\begin{center}
\begin{tabular}{ccc}
\includegraphics[width=.35\textwidth]{./{{img_log_ax_Density-z=16.0-1540}}} & 
\includegraphics[width=.283\textwidth]{./{{img_log_ax_cs-z=16.0-1540}}} &
\includegraphics[width=.283\textwidth]{./{{img_log_ax_vt-z=16.0-1540}}} \\
(a) & (b) & (c) \\
\includegraphics[width=.35\textwidth]{./{{img_log_ax_PGF-z=16.0-1540}}} &  
\includegraphics[width=.295\textwidth]{./{{img_log_ax_Z-z=16.0-1540}}} & 
\includegraphics[width=.295\textwidth]{./{{img_log_ax_PZ-z=16.0-1540}}} \\ 
(d) & (e) & (f) \\
\end{tabular}
\caption{Plots of the gas in a representative star-forming region  at $z=16.0$. Panel (a) depicts gas density, (b) sound speed, (c) turbulent velocity (colors) and velocity in the $x-y$ plane (vector lengths on a linear scale up to $\approx 300 km/s$), (d) pristine gas fraction, (e) average metallicity and (f) average primordial metallicity.   Note that the $\overline Z$ and $\overline Z_{\rm{P}}$ plots are essentially identical indicating that most metals are from Pop III stars. Also, while the average metallicity of much of the gas is non-zero, a high percentage has $P \gtrsim 10^{-1}$, even near the center of the halo.  All plots indicate physical scale and correspond to $20\, {\rm kpc}\, h^{-1}$, comoving. All plots are thin slices in the $z$-plane of the simulation box.}
\label{fig:z16gas}
\end{center}
\end{figure*}

To better understand this small-scale pollution of metals, we examine two representative star-forming regions  in detail.  In Figure~\ref{fig:z16gas}, we show the characteristics of the gas at $z=16,$ in one of the earliest star-forming regions in our simulation. In the center of panel (a) in this figure, we see the overdense region of gas in which star formation and a burst of SNe have recently occurred. In fact, this is the first occurrence of SNe in this halo. We again point out that each SN in our simulation represents the combined action of $\approx 4-20$ SNe, due to our star particle mass range. Panels (b) and (c) depict the impact of the SN burst in terms of the sound speed of the gas (which is $0.15 \, {\rm km/s} \, (T/K)^{1/2}$ ) and the turbulent velocity.  Here we see how SNe heat the surrounding gas to sound speeds of $\gtrsim 200$ km/s, corresponding to  $10^6-10^7$K. These temperatures, while high, are much lower than the initial temperatures at the time at which the SNe occurred in the simulation, because the gas is strongly cooled by adiabatic expansion as the high-pressure regions expand into the intergalactic medium. Thus the radial velocities in figure (c) (vector overlay) are over 300 km/s, resulting in shear strong enough to generate turbulent velocities up to $\approx 500$ km/s. This typically leads to supersonic subgrid turbulent mixing as discussed in \S\ref{sec:local}. The value of the pristine fraction shortly after a burst of SN occurs is set by the mass-density of ejecta relative to the newly polluted cells' mass densities -- as described in \S\ref{sec:pmpg}. Further, while we have demonstrated that the ISM is likely highly turbulent for reasons previously described, we note that in the case of a compressive shock, the subgrid mixing time is proportional to the sound crossing time in the cell: $\Delta x/c_{s}$. This is just another way of saying that the mixing time is proportional to the time it takes the shocked gas, along with the pollutants, to cross the cell. As can be seen in Figure~\ref{fig:z16gas}, in a mostly radially expanding shock, \textit{apparently} dominated by compressive modes, our formulation appropriately models the mixing time, $\tau \propto v_{t}^{-1}$ since in the radial case $v_{t} \propto v_{r}$ where $v_{r}$ is the radial velocity of the shocked gas. This situation only arises in areas of pristine, low density gas that has not been previously stirred by SN or other dynamical effects (e.g. - gravitational shear). However, we say \textit{apparently} here since even a single SN shock expanding into a uniform medium is likely far from purely compressive owing to instabilities in the blast that start within the SN. Such instabilities generate vorticity resulting in clumpy, turbulent ejecta \citep{2007ApJ...661..972P}.

In the lower panels of this figure, we see two regions of gas with $\overline Z \approx 10^{-1}\, Z_{\odot}$ (red areas in panel (e)). These areas also have the highest polluted fractions ($P \lesssim 1\%$, or $f_{\rm pol} \gtrsim 0.99 $) as depicted in panel (d). These areas are found close to the sites of the SNe, in low density gas, and thus they have been turbulently stirred for the longest times.   In addition, the turbulent velocity and sound speed are comparable in these regions-- implying subsonic turbulent mixing, which operates more efficiently than supersonic mixing as discussed above. We also see a very small region of highly pristine gas in the area of high-density gas near the center of the halo. This pocket of gas, with $\overline Z \approx 10^{-3}$, if it were to collapse, is capable of producing star particles with a very high pristine fraction.

\begin{figure*}
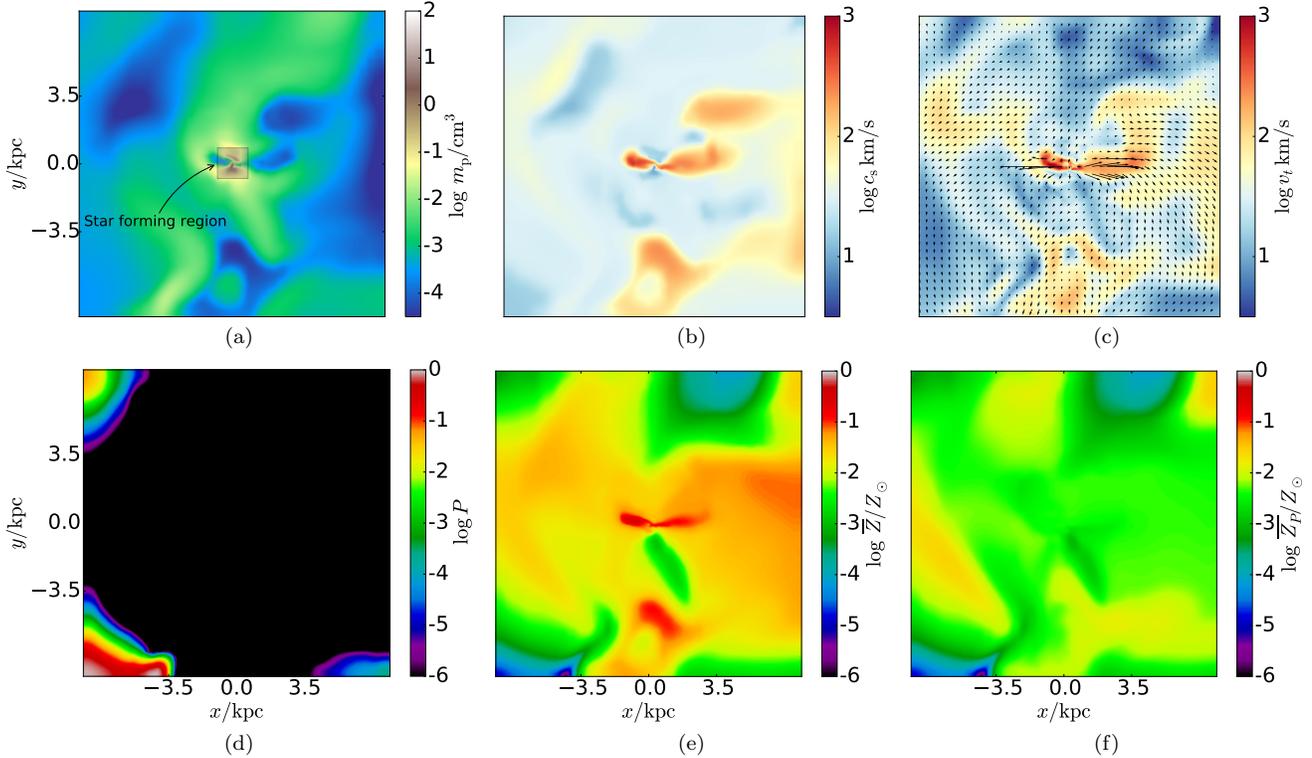

\begin{center}
\setlength\tabcolsep{1mm}
\hspace*{-1cm}
\begin{tabular}{ccc}
 \includegraphics[width=.35\linewidth]{./{{img_log_ax_Density-z=8.0-265793_withBox3}}} & 
 \includegraphics[width=.283\linewidth]{./{{img_log_ax_cs-z=8.0-265793}}} &
 \includegraphics[width=.283\linewidth]{./{{img_log_ax_vt-z=8.0-265793}}} \\ 
 (a) & (b) & (c) \\[2pt]
 \includegraphics[width=.35\linewidth]{./{{img_log_ax_PGF-z=8.0-265793}}} &  
 \includegraphics[width=.295\linewidth]{./{{img_log_ax_Z-z=8.0-265793}}} & 
 \includegraphics[width=.295\linewidth]{./{{img_log_ax_PZ-z=8.0-265793}}} \\
(d) & (e) & (f) \\[2pt]
\end{tabular}
\caption{Plots for $z=8.0$ as in Fig.~\ref{fig:z16gas}, whereas the vectors in panel (c) correspond to a linear scale up to 500 km/s. Comparing the $\overline Z_{\rm{P}}$ and $\overline Z$ plots, $\overline Z_{\rm{P}}$ is $\approx 1$ dex down over most regions indicating pollution by a second-generation of stars. All plots indicate physical scale at the redshift indicated and correspond to $100\, {\rm kpc}\, {\rm h}^{-1}$, comoving. The region of star formation called out in the density plot is used in Fig.~\ref{fig:z8breakout}.}
\label{fig:z8gas}
\end{center}
\end{figure*}

Moving farther out from the center, we see that the average metallicity of the gas falls off relatively slowly, indicating that polluted material has been carried most of the way across this region.  However, the radial increase in  the pristine fraction is much steeper, moving to $\gtrsim 10\%$ in $\overline Z \lesssim 10^{-2.5} Z_\odot$ areas and increasing to  $\approx 100\%$  in areas where $\overline Z \lesssim 10^{-4} Z_\odot$. This unmixed gas  occurs in the regions in which the turbulent velocity drops sharply, illustrating the correlation between turbulent mixing, the kinetic energy, and the sound speed in the gas. Finally, we note that the vast majority of metals are from primordial stars, indicated by the near identical plots for $\overline Z_{\rm P}$ and $\overline Z$.

Figure~\ref{fig:z8gas} depicts the gas in a $100\, {\rm kpc}\, h^{-1}$ comoving star-forming region at $z=8$. Unlike in the early halo case shown in Figure \ref{fig:z16gas}, most of the gas  in this case is thoroughly mixed. The metallicity plot shows that the gas has been enriched to at least $\overline Z \geq 10^{-3}\, Z_{\odot}$ throughout most of this slice. Looking at the sound speed and turbulent velocity plots we note that, in general, there are few regions where $v_{\rm t} \gg c_{\rm s}$, indicating that SN energy has mostly been disbursed: another indication that there has been sufficient time for mixing to occur. On the other hand, there remain pockets of gas around the edge of this slice in which the overall metallicity is relatively high, $\overline Z  \gtrsim 10^{-3} Z_{\odot}$, but where $P \gtrsim 10^{-2}$. These regions of incomplete mixing correlated with regions of lower density. By comparing $\overline Z$ to $\overline Z_{P}$,  we note that, roughly, only 10\% of the metals in this region are primordial indicating thorough pollution by a second-generation of SN. Interestingly however, there is one region just below the central star-forming region with $\overline Z \approx 10^{-3} Z_{\odot}$ with densities of $10^{-1}\,m_{\rm p}\, {\rm cm}^{-3} <\rho \leq 10^{-2.5}\,m_{\rm p}\, {\rm cm}^{-3} $ where we find $\overline Z \approx \overline Z_{P},$ indicating that most of the metals here are likely from first generation ejecta. Should subsequent star formation occur in this moderately dense region we would expect a population of stars with a high fraction of primordial metals. Even in this seemingly well polluted region our two new scalars have allowed us to paint a more nuanced picture of metal-enrichment of the gas, that will lead to important observable differences for the resulting stellar populations.

\subsection{Stellar Populations}\label{sec:stars}

\begin{figure*}
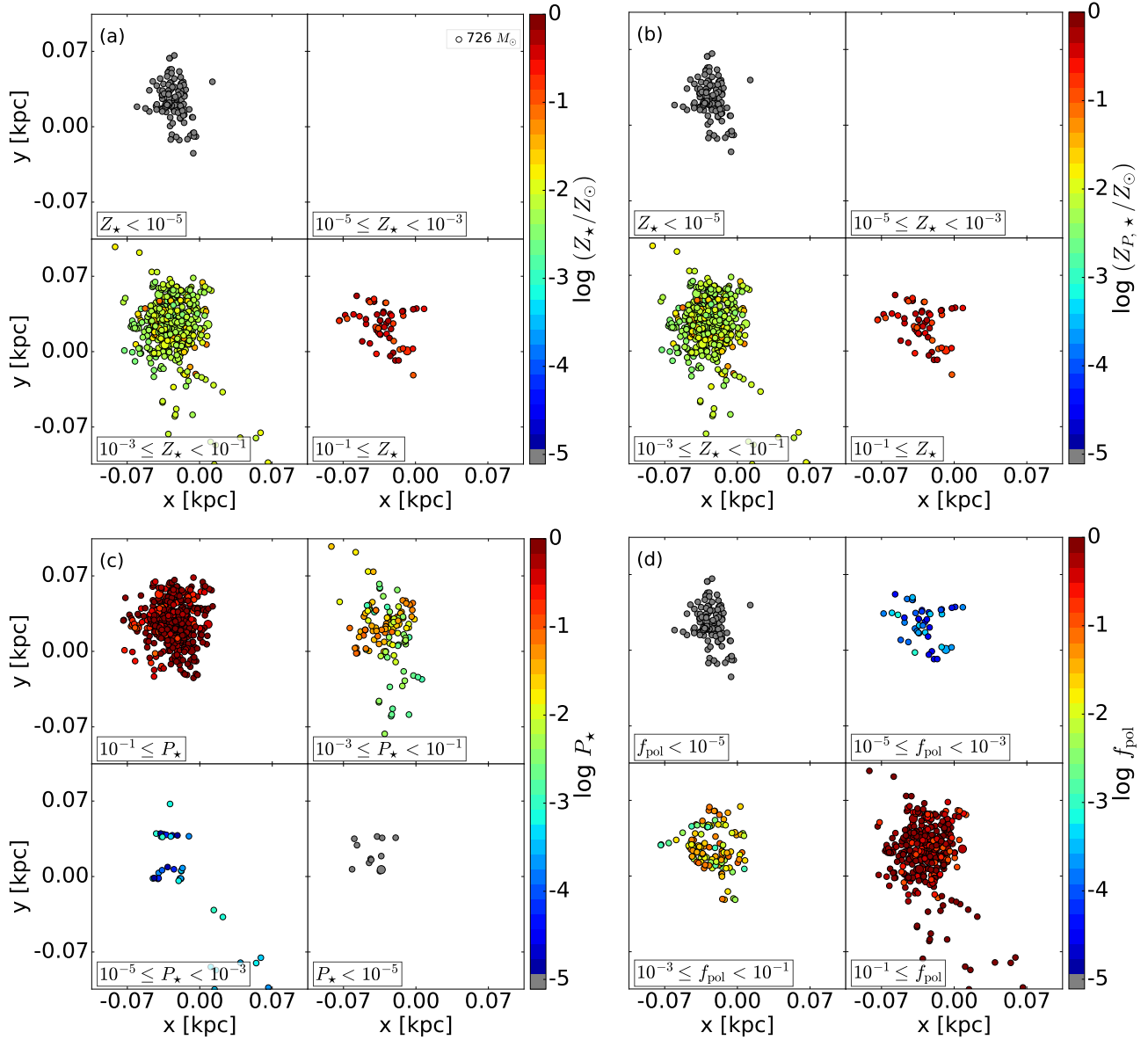

\begin{center}
\hspace*{-0.5cm}
\begin{tabular}{cc}
 \includegraphics[width=0.5\textwidth]{./{{SP_Z_locs_Zcut_z=16.0-770}}} & 
 \includegraphics[width=0.434\textwidth]{./{{SP_PZ_locs_Zcut_z=16.0-770}}} \\ 
 \includegraphics[width=0.5\textwidth]{./{{SP_PF_locs_PFcut_z=16.0-770}}} & 
 \includegraphics[width=0.434\textwidth]{./{{SP_fpol_locs_fpolcut_z=16.0-770}}} \\ 
\end{tabular}
\caption{Properties of the stars formed in a halo at $z=16$. Each group of 4 plots depicts star particle locations in physical ${\rm kpc}$, 2.5 kpc $h^{-1}$ comoving. Dot size indicates star particle mass in $M_{\odot}$. Panels (a) \& (b) depict corrected metallicity \& primordial metallicity binned in 4 metallicity subranges.  These are nearly identical, indicating that almost all (99\%) of the metals in this halo are primordial (from Pop III SN).  Panels (c) \& (d) depict star particle pristine fraction and the polluted fraction binned in 4 subranges.  While many of these star particles have metallicities in the range $10^{-3} Z_\odot \leq Z_{\star} < 10^{-1} Z_\odot $, a significant fraction of the stars they represent are Population III (panel (c), $10^{-1}  \le P_{\star}$).  }
\label{fig:z16sp}
\end{center}
\end{figure*}

\begin{figure*}
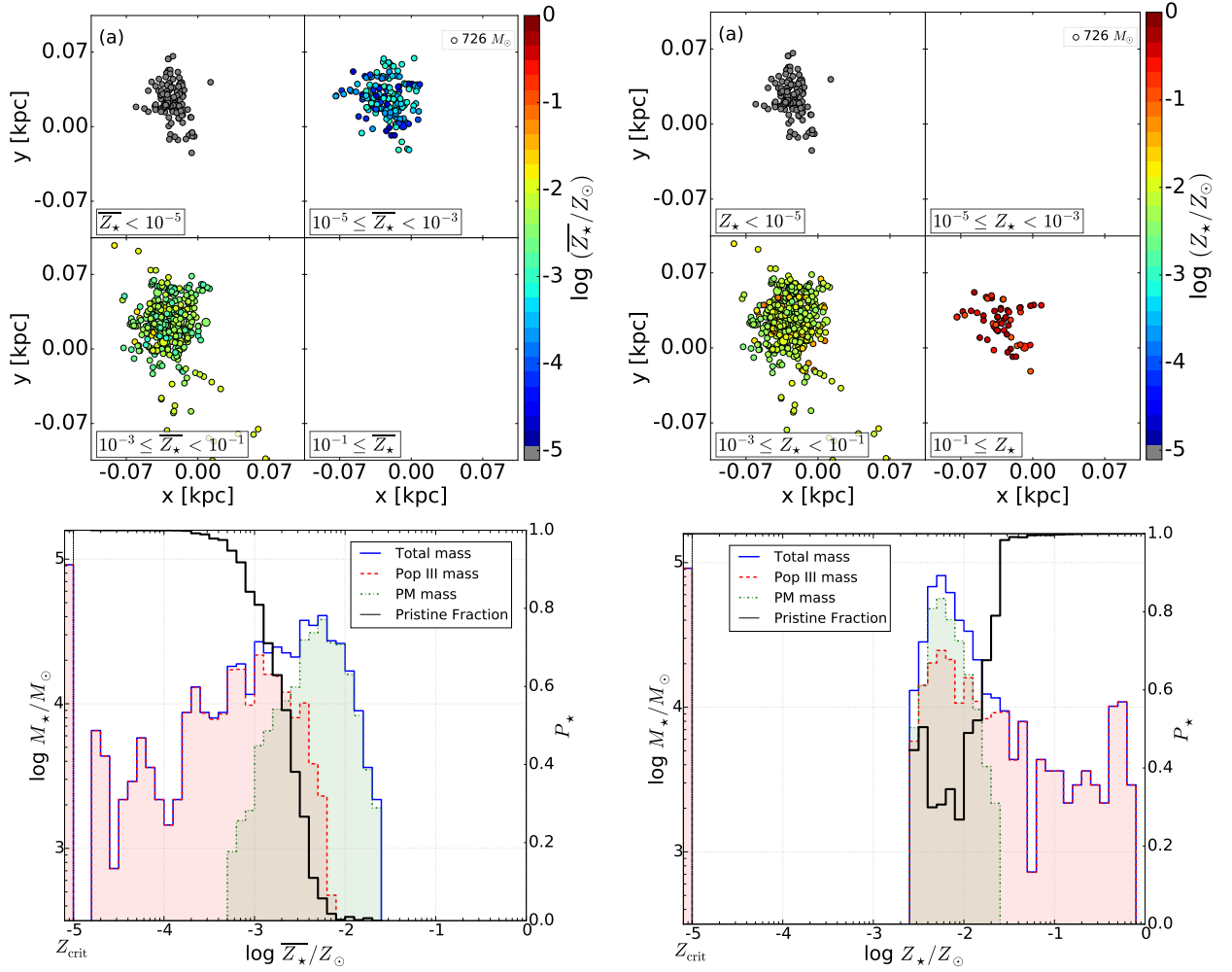

\begin{center}
\begin{tabular}{cc}
 \includegraphics[width=0.45\textwidth]{./{{SP_Z_locs_Zcut_uncorr_z=16.0-770}}} & 
 \includegraphics[width=0.45\textwidth]{./{{SP_Z_locs_Zcut_z=16.0-770}}} \\ 
 \includegraphics[width=0.42\textwidth]{./{{Hist_HaloData_z=16.0SFRegion_dualAx}}} &
 \includegraphics[width=0.42\textwidth]{./{{Hist_HaloData_z=16.0SFRegion_fpol_dualAx}}}\\ 
\end{tabular}
\caption{Histograms depicting star particle pristine and primordial mass binned in the particles' associated average (left) or corrected (right) metallicity for a star-forming region at $z=16$ using 50 logarithmic metallicity bins from $10^{-5}\, Z_\odot$ to solar. We have binned all classical Pop III stars ($Z_{\star}$ or $\overline Z_{\star} < Z_{\rm crit}$) just below the bin at $Z_{\rm crit} = 10^{-5} Z_\odot$. The histograms' right axes depict the particles' pristine fraction in each metallicity bin. The Pop III mass in each metallicity bin corresponds to $P_{\star}\times M_{\star}$ (red) and the primordial metal (PM) mass is $f_{\rm pol}\times M_{\star}\times Z_{P \star}/Z_{\star}$ (green). Comparing the two panels, we see that the relatively small $f_{\rm pol}$ of gas in this halo significantly enhances the metallicity of the star particles with $\overline Z_{\star} \lesssim 10^{-2.5} Z_\odot.$ In fact, 99.6\% of star particle mass with $Z_{\star} \gtrsim 10^{-1.6} Z_\odot$ (right panel) represents Pop III stars. Overall, 98.6\% of stellar metals are primordial and these polluted stars have metallicities $10^{-2.8} Z_\odot < Z_{\star} < 10^{-1.6} Z_\odot$.}
\label{fig:z16hists}
\end{center}
\end{figure*}

Figure~\ref{fig:z16sp} shows several properties of stars formed in a representative minihalo at $z=16$. Note that the metallicities in this figure are corrected to correspond to the metallicities of the subset of enriched stars within the star particle, $Z_{\star} = \overline Z_{\star}  f_{\rm pol}^{-1}$, as described in \S2. 
Here we have used the upper bound for the correction to the metallicity since we are at high redshift and the gas in and around the halos is in the process of being enriched by a group of SN.
As the star particles in the uppermost left panel of subplot (a) in this figure have $Z_{\star} \leq Z_{\rm crit},$ even after correcting for $f_{\rm pol}$, they represent stellar mass composed of pristine material, labeled as `classical' Pop III stars in Figure~\ref{fig:sfr}.  Such particles make up about 17\% of the stellar mass of this early halo, as compared to the majority of the star particles which fall into the metallicity bin $10^{-3} Z_\odot < Z_{\star} \leq 10^{-1} Z_\odot.$ Interestingly, correcting for the polluted fraction of gas results in no stars in the metallicity bin $10^{-5}  Z_\odot < Z_{\star} \leq 10^{-3} Z_\odot.$ This indicates that the polluted fraction of the cell must have been relatively small when ejecta in this range polluted star-forming cells, resulting in enhanced metallicities ($ Z_{\star}  > 10^{-3}\,  Z_\odot$) due to the concentration of the metals.  This can be easily seen in the top panels of Figure~\ref{fig:z16hists}. Examining the pristine and polluted fraction plots, its apparent that many of the higher $Z_{\star}$ star particles reach these high corrected metallicities only because they not only have pristine fractions, $P_{\star}$ greater than 10\%, but  polluted fractions $f_{\rm pol} = 1-P_{\star}$ less than 10\%, implying $P_{\star} > 0.9$; meaning that the majority of the gas from which they were formed was unmixed. In fact if we include the pristine fractions of all star particles, the Pop III fraction for this halo grows to 57\%, an increase of a factor of nearly 3.4. However, we note  several star particles in panel (a),  $10^{-3}  Z_\odot < Z_{\star} \leq 10^{-1} Z_\odot$, to the lower-right of the central concentration of particles, that do not appear in the pristine fraction plot for $P_{\star} > 10^{-1}$.  These star particles formed at a later time in gas that was more thoroughly mixed and represent a relatively small population of almost all Pop II stars with $10^{-5}< P_{\star} \le 10^{-3}$. Additionally, by comparing the spatial distribution of stars as a function of metallicity, $Z_{\star},$ to the distribution of stars as a function of primordial metallicity, $Z_{\rm P,\star},$ we can see that almost all of the metals in the enriched stars within this halo are produced by Pop III stars.

\begin{figure*}
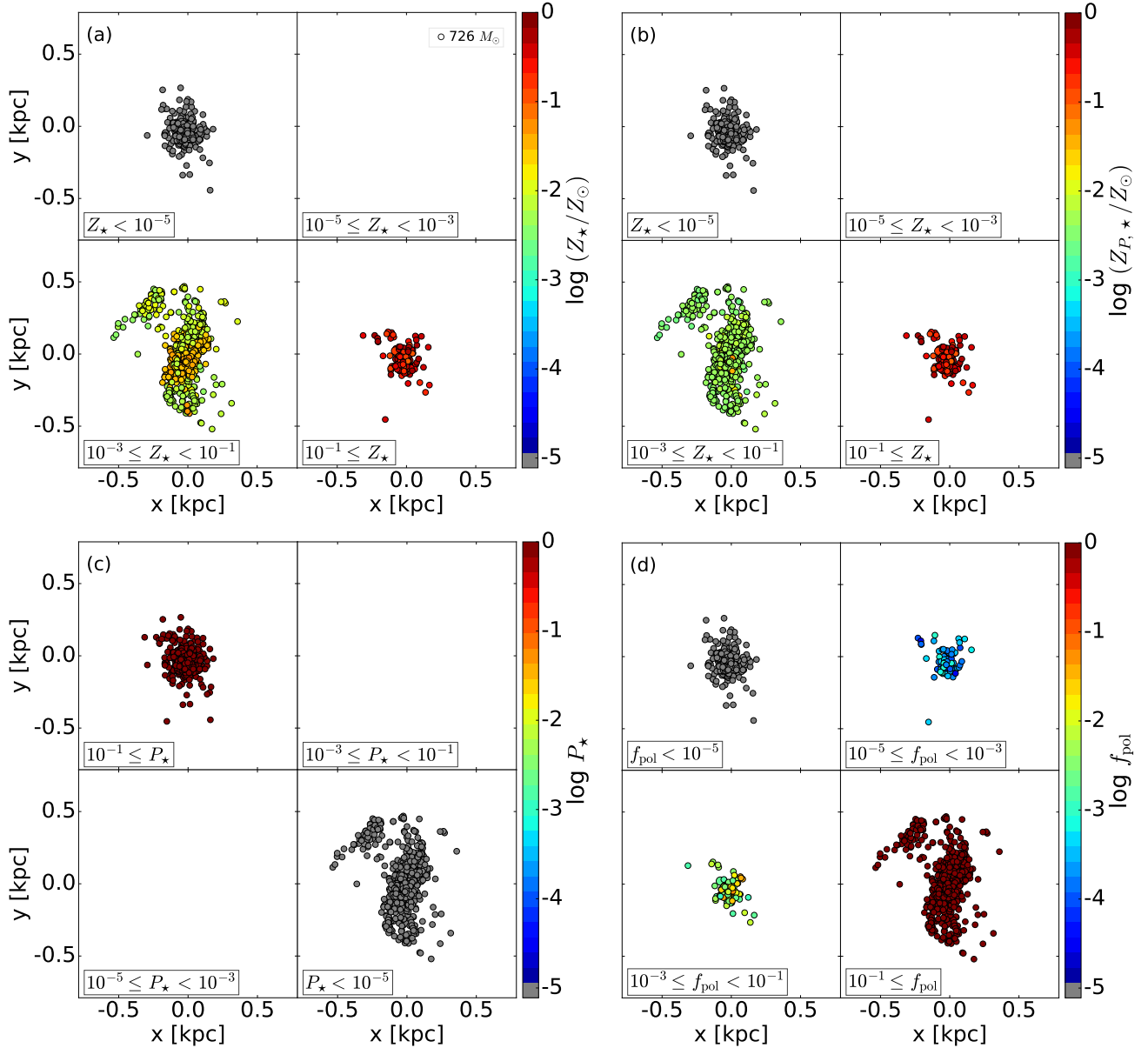

\begin{center}
\begin{tabular}{cc}
 \includegraphics[width=0.5\textwidth]{./{{SP_Z_locs_Zcut_z=08.0-265793}}} & 
 \includegraphics[width=0.442\textwidth]{./{{SP_PZ_locs_Zcut_z=08.0-265793}}} \\ 
 \includegraphics[width=0.5\textwidth]{./{{SP_PF_locs_PFcut_z=08.0-265793}}} & 
 \includegraphics[width=0.442\textwidth]{./{{SP_fpol_locs_fpolcut_z=08.0-265793}}} \\ 
\end{tabular}
\caption{Halo at $z=8$. Panels as in Fig.~\ref{fig:z16sp}. For this halo we note the $PM$ is down $\approx 1$ dex from the overall metallicity of the stars in the range $10^{-3} Z_\odot \le Z_{\star} < 10^{-1} Z_\odot $ (panels (a) and (b)). However, there is still a relatively large fraction of pristine stars in this halo as discussed in the text. The comoving scale is 9.5 kpc $h^{-1}$ with the axis labeled in physical kpc.}
\label{fig:z8breakout}
\end{center}
\end{figure*}

These conclusions are supported by Figure \ref{fig:z16hists}, which shows the total, Pop III, and primordial stellar masses ($f_{\rm pol}\times M_{\star}\times \frac{Z_{P \star}}{Z_\star}$) binned according to the star particles' corrected metallicity (right side). For comparison, we also include an uncorrected metallicity histogram to show the effects of the $f_{\rm pol}$ correction to $\overline Z_{\star} $.  While there are a fair number of star particles with average metallicities in the range $10^{-5} Z_\odot < \overline Z_{\star} \lesssim 10^{-2.6} Z_\odot,$ the lowest nonzero corrected metallicities are $\approx 10^{-3.1} Z_\odot.$  In fact, after correcting for the polluted fraction, these extremely low average metallicity star particles are amongst the star particles with the highest corrected metallicities, such that star particles with $Z_{\star} \gtrsim 10^{-1.6} Z_\odot$ are over 99.6\% pristine. These are the star particles that form at the edges of the extent of enriched material, still within the central star-forming region, in which the average gas metallicity is low but very little mixing has occurred, such that the metals are concentrated into just a few stars.

As mentioned above, we have not modeled radiative feedback in our simulation and our example halo at z=16 formed classical Pop III stars in two successive yet closely spaced ($\Delta t < 7$ Myr) waves. Even though radiative feedback from these stars would likely lower densities and suppress further central star formation \citep{2012AIPC.1480..123W}, it is also possible for feedback to trigger nearby collapse in dense clumps in the pristine gas. Our simulation indicates subsequent significant star formation in this halo takes place on a time scale of $< 20$ Myr, such that star particles still have relatively high pristine fraction, $P > 0.1$.  While we feel these results are representative it will be interesting to model radiative feedback and quantify the differences.

Moving to $10^{-2.8} Z_\odot <  Z_{\star} \lesssim 10^{-1.6} Z_\odot$ star particles formed in more vigorously stirred material, we find that 38\% of this material is composed of Pop III stars. In fact, only when one also includes star particles with average metallicities up to $10^{-1.6} Z_\odot$ does one arrive at a 57\% mass fraction of Pop III stars in this object.  Finally, from the primordial stellar mass histograms we can see that the majority of enriched stars are found in $10^{-2.5} Z_\odot < \overline  Z_{\star}   \lesssim 10^{-1.5} Z_\odot$ particles, and that in these particles, essentially all the metals are generated by Pop III stars.

In Figure \ref{fig:z8breakout}, we  turn our attention to the properties of the stars formed in a halo at $z=8.$   Unlike Figure \ref{fig:z16sp}, which shows a different $z=16$ region than shown in the gas plots in Fig. \ref{fig:z16gas}, Fig. \ref{fig:z8breakout} shows the stellar distribution in the same region at the center of the $z=8$ region depicted in Figure \ref{fig:z8gas}. This halo is older than the $z=16$ halo and has formed stars over a longer period, resulting in more metals and significantly more mixing.  However, while the more extended history shifts the average metallicity to somewhat higher values than in the $z=16$ case, they still share many features.  In both halos, the largest collection of star particles occurs in the $10^{-3}  Z_\odot < Z_{\star} \leq 10^{-1} Z_\odot$ range, in both halos there are large number of star particles with $f_{\rm pol} < 0.1$, and in both halos a significant number of Pop III stars are formed.

\begin{figure}[h]
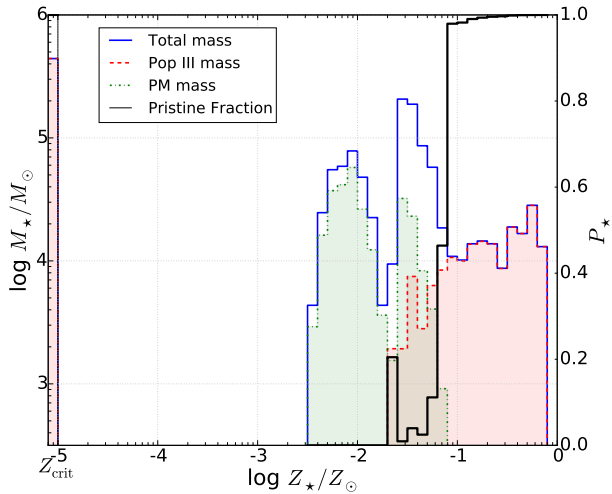

\begin{center}
 \includegraphics[width=0.45\textwidth]{./{{Hist_HaloData_z=08.0SFRegion_fpol_20kpc_dualAx}}} 
\caption{Metallicity histogram for the halo at $z=8$ binned as in Fig.~\ref{fig:z16hists} with pristine fraction on the right axis. We note three distinct stellar populations: one with $10^{-2.6} < Z_{\star} \leq 10^{-1.8}$ consisting mostly of older stars with a high fraction (67\%) of primordial metals. Another population of stars between $10^{-1.8} < Z_{\star} \lesssim 10^{-1.1}$ has a much lower fraction of $PM$, 13\%, and composed mostly (95\%) of Pop II stars. Finally, 99.5\% of the stellar mass with $Z_{\star} \geq 10^{-1.1}$ represent Pop III stars.}
\label{fig:z8hist}
\end{center}
\end{figure}

On the other hand, unlike the higher redshift case, second-generation metals are present in the $z=8$ halo, as can be seen by comparing the total metallicity in subplot (a) to the primordial metallicity in subplot (b).  This shows that the primordial metals are down by a factor of $\approx 3$ from the $Z_{\star}$ levels.  And while there is a large fraction of star particles with $P_{\star} > 0.1$, there is also now a comparable fraction with $P_{\star} \leq 10^{-5}$ indicating a large fraction of star particles fully polluted with metals.

\begin{deluxetable*}{cccccc}
\tabletypesize{\footnotesize} 
\tablecolumns{6} 
\setlength{\tabcolsep}{1pt}
\tablecaption{Halo characteristics} 
\tablehead{
\colhead{} & \colhead{Total} & \colhead{Pop III} & \colhead{$\underline{\rm Classic\; Pop III}$} & &  \\
\colhead{Redshift} & \colhead{$\nicefrac{M}{M_{\odot}}$} & \colhead{$\nicefrac{M}{M_{\odot}}$} & \colhead{Pop III} & \colhead{$\langle{Z_{\star}\rangle}$\tablenotemark{a}} & \colhead{$\langle \nicefrac{Z_{\rm P,\star}}{Z_{\star}}\rangle$\tablenotemark{b}} }
\startdata 
16 & $5.35\times 10^{5}$ & $3.04\times 10^{5}$ & $0.300$ & $3.07\times 10^{-3}$ & $0.988$  \\
8 & $1.45\times 10^{6}$ & $6.19\times 10^{5}$ & $0.714$ & $1.53\times 10^{-2}$ & $0.322$  
\enddata \label{table:halostats}
\tablenotetext{a}{Mass-weighted average metallicity for all stars in the halo.}
\tablenotetext{b}{Mass-weighted average primordial metal fraction for polluted stars in the halo.}
\tablenotetext{ }{}
\end{deluxetable*}

\begin{figure*}
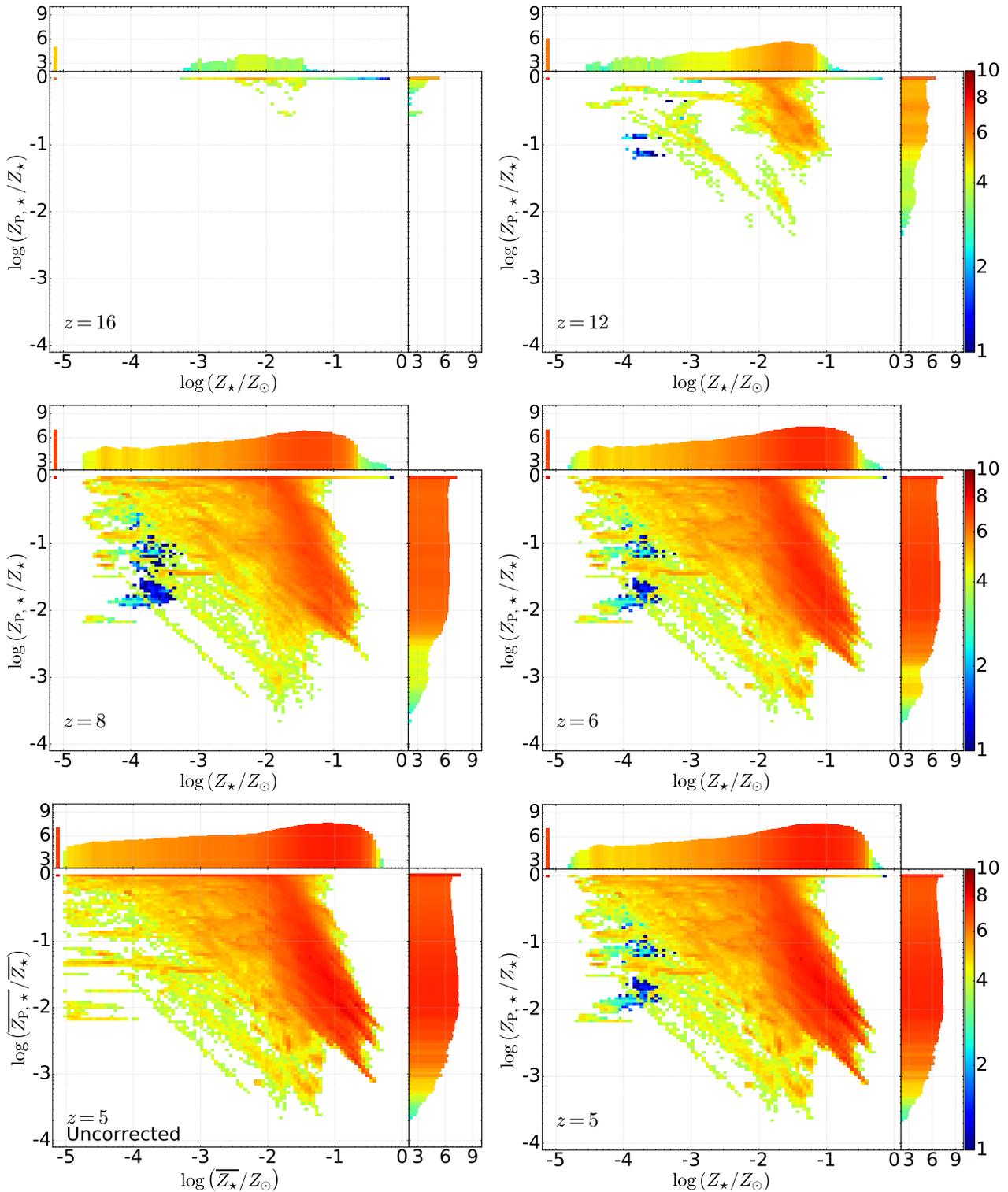

\begin{center}
\setlength{\tabcolsep}{1mm}
\hspace*{-1cm}
\begin{tabular}{cc}
 \includegraphics[width=0.45\textwidth]{./{{Z_fpol-vs-Z_P-MassHistLogFullNorm-z_16.00}}} &   
 \includegraphics[width=0.474\textwidth]{./{{Z_fpol-vs-Z_P-MassHistLogFullNorm-z_12.00}}} \\ 
 \includegraphics[width=0.45\textwidth]{./{{Z_fpol-vs-Z_P-MassHistLogFullNorm-z_08.00}}} &
 \includegraphics[width=0.474\textwidth]{./{{Z_fpol-vs-Z_P-MassHistLogFullNorm-z_06.00}}} \\  
 \includegraphics[width=0.45\textwidth]{./{{Z-vs-Z_P-MassHistLogFullNorm-z_05.00}}} &
 \includegraphics[width=0.474\textwidth]{./{{Z_fpol-vs-Z_P-MassHistLogFullNorm-z_05.00}}}  
\end{tabular}
\caption{Evolution of the joint mass-weighted PDFs for our star particles. Units are $M/M_{\odot}$ normalized to the (logarithmic) bin size and comoving volume of the simulation, all metal-free star particles are represented at [-5,0].  There is a clear trend toward smaller primordial fractions, $Z_{\rm P,\star}/ Z_{\star} < 1$, in higher metallicity stars, $Z_{\star} \ge 10^{-3}\, Z_\odot$.  However purely primordial star particles  $Z_{\rm P,\star} =  Z_\star$ are found over a range of metallicities ($10^{-3} Z_\odot \lesssim Z_{\star} \lesssim 10^{-0.5} Z_\odot$). For comparison, we also include a PDF for the uncorrected metallicities (lower left panel), which is dramatically different at low metallicities.  }
\label{fig:zpol}
\end{center}
\end{figure*}

Figure~\ref{fig:z8hist} depicts the metallicity distribution for the star particles in this minihalo. As visible in the scatter plots, a much larger fraction of stars in the range $10^{-1.8} Z_\odot < Z_{\star} \lesssim 10^{-1.1} Z_\odot$ are polluted (95\%) and the fraction of primordial metals, 0.13, is much lower than in the earlier halo.  However, looking at stars in the range $10^{-2.7} Z_\odot < Z_{\star} \leq 10^{-1.8} Z_\odot$ we again see a population of Pop II stars with a high fraction of primordial metals. These most likely represent an older population that was mostly polluted with Pop III material, which are likely good CEMP-no candidates. We also note another large Pop III group of stars born in largely unmixed gas with $Z \geq 10^{-1.1} Z_\odot$. In fact, 99\% of the stars in these star particles are pristine.  For comparison, Table~\ref{table:halostats} captures some of the relevant characteristics of the two halos we have been discussing. Here, angle brackets around the metallicity indicates a mass-weighted average taken over all stars in the halo. 

Finally, we examine the evolution of the metallicity of our star particles in aggregate. Figure \ref{fig:zpol} depicts the evolution of the mass-weighted probability density function (PDF) for our star particles, with the primordial metal fraction, $Z_{\rm P, \star}/Z_{\star}$ plotted against the corrected metallicity, $Z_{\star}$.  As expected, as metallicity increases the fraction of primordial metals decreases (at every epoch) giving the plots their characteristic negative slope.  By redshift 5,  the majority of the stars have metallicities in the range $10^{-1.8} Z_{\odot} < Z_{\star} < 10^{-0.5} Z_{\odot}$.  As we move to extremely metal poor populations, $Z_{\star} < 10^{-3}$, the negative slope of our PDF indicates that the fraction of primordial metals increases as overall metallicity decreases such that the most stars have $Z_{\rm{P}}/Z_{\star} > 10^{-1}$.  These Population II stars are likely CEMP-no candidates.  Furthermore the distribution of stars with $Z_{\star} < 10^{-1.5}\,Z_{\odot}$, remains roughly constant below $\textit{z} = 6$ indicating that most of the low-metallicity stars, including those with a high fraction of primordial metals, have been formed before this epoch. 

As in previous diagrams, we have used the upper bound for the corrected metallicity to make these plots. For comparison, the lower left plot, annotated `Uncorrected', shows the effects of not making the polluted fraction correction to the metallicity at $z=5.$  The $f_{\rm pol}$ correction is greatest for star particles with $\overline Z_{\star} \lesssim 10^{-3.5}$ since these were contaminated by material with both a low average metallicity and a small $f_{\rm pol}$. Without this correction we would predict a larger number of stars with metallicities near $Z_{\rm crit},$ while the corrected results show very few stars enhanced to values below $10^{-4} Z_\odot.$  The correction also shifts the population of purely primordial star particles, $Z_{\rm{P}}/Z_{\star} = 1$ at $z=5$, away from very low metallicities, concentrating them in the range $10^{-3} Z_{\odot} < Z_{\star} < 10^{-1} Z_{\odot}$.  While stars in this metallicity range are not extremely metal poor, they do possess a large fraction of primordial metals and may represent an important iron-poor stellar population. 

\begin{figure*}
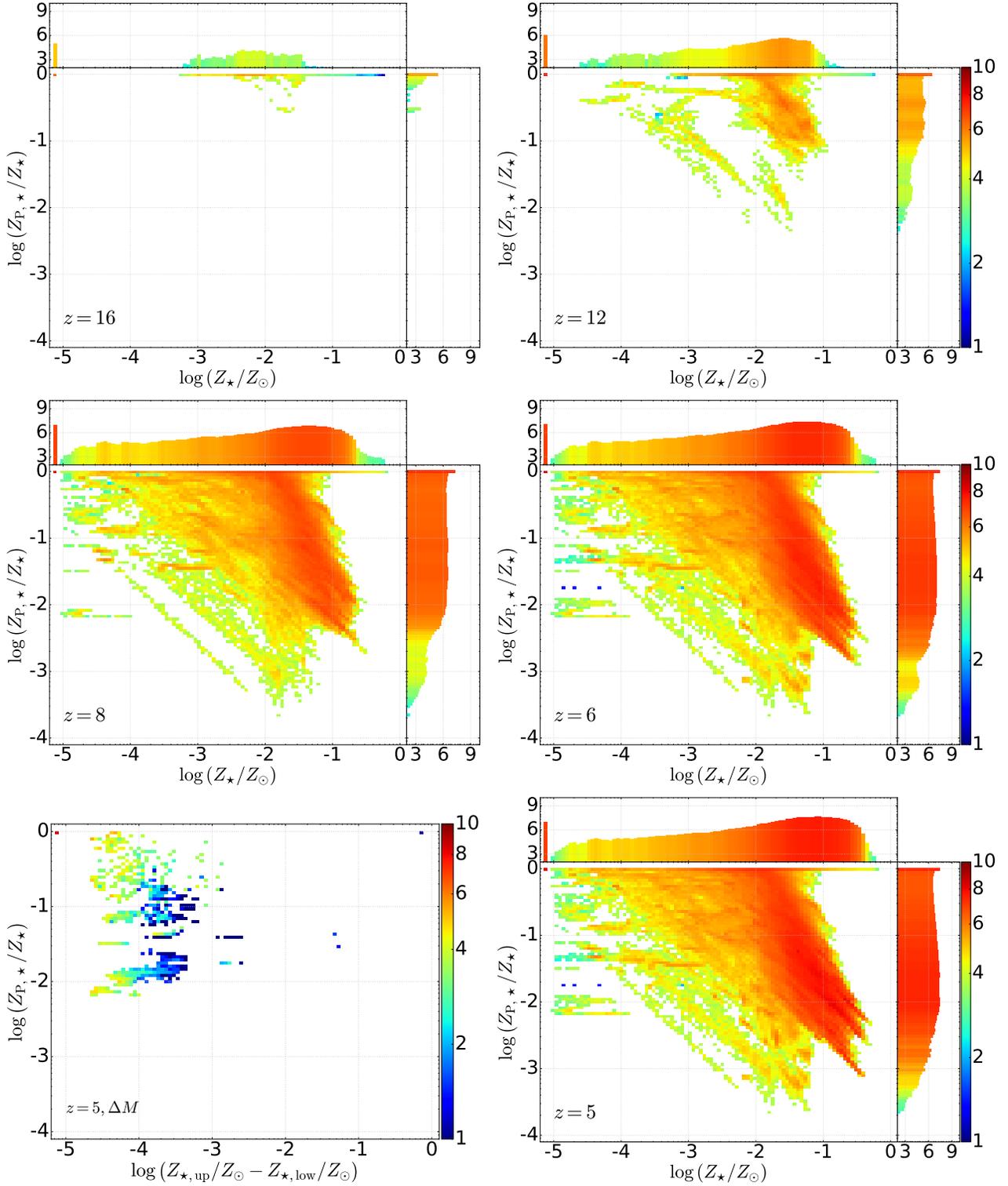

\begin{center}
\setlength{\tabcolsep}{1mm}
\hspace*{-1cm}
\begin{tabular}{cc}
 \includegraphics[width=0.45\textwidth]{./{{Z_fpol-vs-Z_P-MassHistLogFullNorm-z_16.00_lowerBound}}} &   
 \includegraphics[width=0.474\textwidth]{./{{Z_fpol-vs-Z_P-MassHistLogFullNorm-z_12.00_lowerBound}}} \\ 
 \includegraphics[width=0.45\textwidth]{./{{Z_fpol-vs-Z_P-MassHistLogFullNorm-z_08.00_lowerBound}}} &
 \includegraphics[width=0.474\textwidth]{./{{Z_fpol-vs-Z_P-MassHistLogFullNorm-z_06.00_lowerBound}}} \\  
 \includegraphics[width=0.45\textwidth]{./{{Z_fpol-Z-xaxis_diff-z_05.00_diff}}} &
 \includegraphics[width=0.474\textwidth]{./{{Z_fpol-vs-Z_P-MassHistLogFullNorm-z_05.00_lowerBound}}}  
\end{tabular}
\caption{Joint mass-weighted PDFs for our star particles using the lower bound on the metallicity correction to correct star particle metallicity. Units are as in the previous figure.  Here we note more mass in the very low ($Z_{\star} \lesssim 10^{-4} Z_\odot$) bins since the lower bound correction assumes that the maximal amount of metals are tied up in Pop III stars ($Z_{\star} \approx Z_{\rm crit}$). The lower-left panel depicts the difference between the upper-bound and lower-bound metallicities for all $z=5$ star particles. }
\label{fig:zpollow}
\end{center}
\end{figure*}

For comparison, we include PDF plots for the lower bound on the correction to $Z_{\star}$. As can be seen in Figure \ref{fig:zpollow}, if we assume that the maximum amount of metals in these star particles is distributed in the Pop III stars, at a level $Z_{\star} \lesssim Z_{\rm crit}$, we reduce the enhancement of the lowest metallicity stars. This is clearly depicted in lower-left panel where we have differenced the masses in the metallicity bins between the upper and lower bounds for star particles at $z=5.0$.

\subsection{Chemical Evolution}\label{sec:chem}

\begin{figure*}
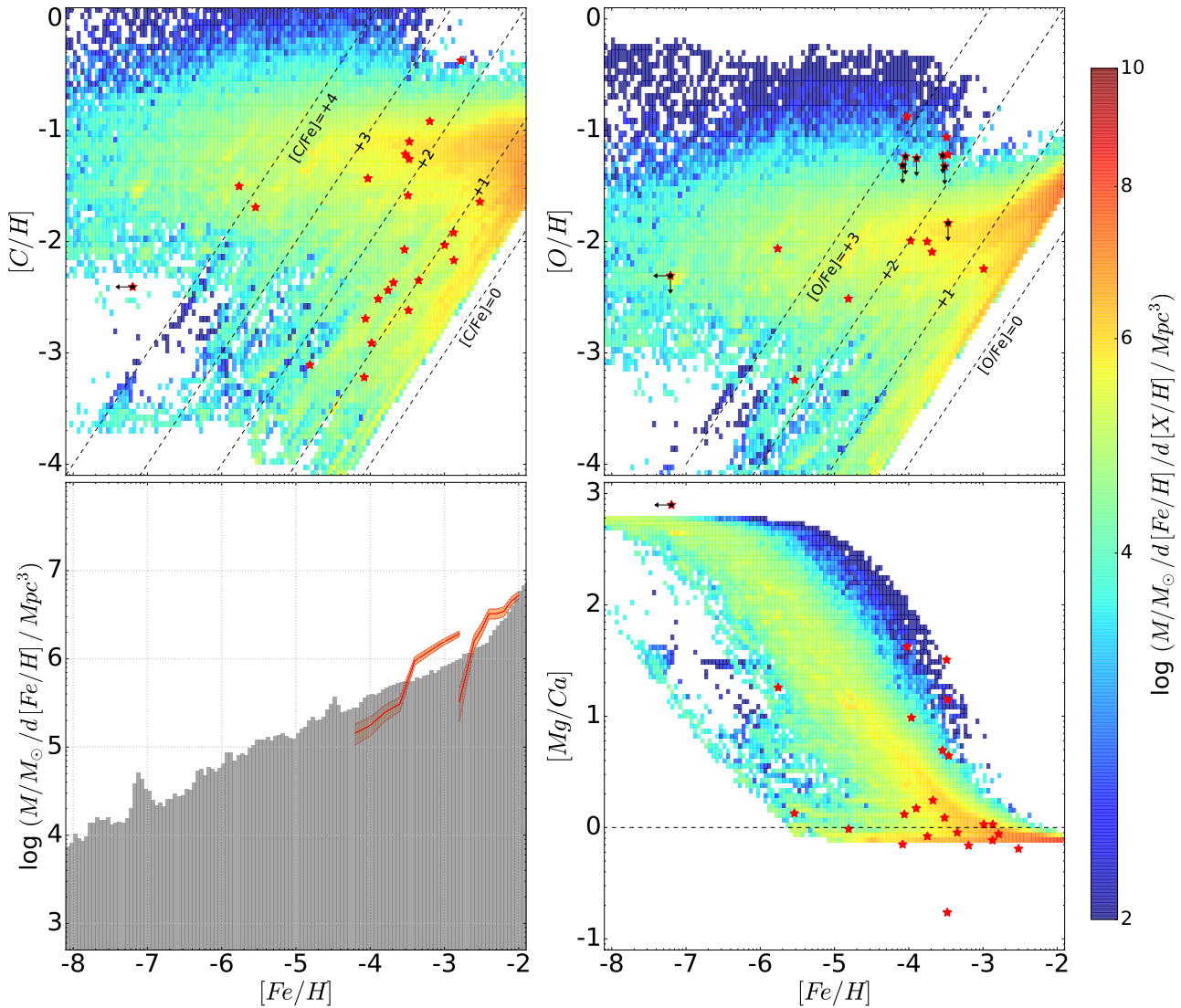

\begin{center}
\setlength{\floatsep}{1pt plus 1.0pt minus 2.0pt}
 \includegraphics[width=.95\textwidth]{./{{metals_fpol}}} \\ 
\caption{Joint PDFs depicting the mass-weighted probabilities for the chemical abundances of [C/H], [O/H] and [Mg/Ca] as a function of [Fe/H] for stars in our simulation. The histogram (bottom left) bins our stars' [Fe/H] abundance and captures our MDF. The histogram is overlaid with observational MDFs for a population of metal poor stars selected from \cite{2016arXiv160706336Y} (lower MDF) and \cite{2013ApJ...763...65A} (upper MDF). Our PDFs correlate well with the observations of CEMP-no stars from \cite{2014Natur.506..463K} (red stars). Note that our MDF histogram includes all stars in our simulation, while the observational data focus on metal poor and CEMP(-no) stars.}
\label{fig:chem}
\end{center}
\end{figure*}

The two types of metallicity associated with our star particles, $Z_{\star}$ and $Z_{\rm P, \star}$, provide us with the information needed to more accurately model their chemical make-up, and compare them with observations. Motivated by the lack of $Z \leq 10 ^{-1.5} Z_\odot$ Pop II star formation below $z=6,$ we assume that at $10^{-1.5}\,Z_{\odot}$ the final distribution of enriched stars in our simulation can be compared with the distribution of metal-poor halo stars, and we model the abundances of our stars as the convolution of the ejecta of a single type of Pop III SN (for $Z_{\rm P, \star}$) and a single-set of abundances representative of second-generation metals (for $Z_{\star}$) at $z=5$. While a more heterogeneous model for Pop III SN ejecta would be more physical \citep{2002ApJ...567..532H, 2014ApJ...792L..32I},  it would also introduce additional parameters into the problem, and thus we focus in this first paper on a single representative model for this study. Specifically, we model the metallicity produced by primordial stars as the yield from $60\, M_{\odot}$ Pop III SN \citep{hegerwebsite}. This very iron poor and carbon enhanced ejecta is a best-fit for the CEMP-no star SMSS J031300.36-670839.3 studied by \cite{2014Natur.506..463K} and is representative of core-collapse SN yields in the  $25 M_{\odot} \leq M_{\star} \leq 140 M_{\odot}$ range in which metal yields are believed to be carbon-enhanced and iron poor \citep{2002ApJ...567..532H}. The composition typical of $z=5$ was determined by a chemical evolution model run to that epoch by \cite{Timmes}. We capture the characteristics of these two types of material in Table~\ref{table:chem}.

\begin{deluxetable}{ccc} 
\tabletypesize{\footnotesize} 
\tablecolumns{3} 
\setlength{\tabcolsep}{1mm}
\tablewidth{\columnwidth} 
\tablecaption{Mass fractions of metals \label{table:chem}} 
\tablehead{
\colhead{} & \colhead{${X/Z}$} & \colhead{$X/Z_P$} \\
\colhead{Element}  & \colhead{1 Gy} & \colhead{60 $M_{\odot}$ Pop III SNe}
} 
\startdata 
 C & 1.68 $\times 10^{-1}$ & 7.11 $\times 10^{-1}$ \\
 O & 5.29 $\times 10^{-1}$ & 2.73 $\times 10^{-1}$ \\
Mg & 2.49 $\times 10^{-2}$ & 9.56 $\times 10^{-4}$ \\
Ca & 3.00 $\times 10^{-3}$ & 1.43 $\times 10^{-7}$ \\
Fe & 5.39 $\times 10^{-2}$ & 2.64 $\times 10^{-12}$
\enddata 
\tablenotetext{}{The mass fractions of metals for selected elements used to model the normal and primordial metallicity of star particles in our simulation. Data for gas typical of $1\,Gyr$ post BB provided by \cite{Timmes}. Data for $60 M_{\odot}$ Pop III SN provided by \cite{hegerwebsite}.}
\end{deluxetable}

Using the $60 M_{\odot}$ Pop III SN yields, we model [C/H], [O/H] and [Mg/Ca] to [Fe/H] ratios for stars in our simulation as \begin{equation}\label{ref:metalConvo}
\begin{aligned}
Z^{i}_{\star} = \left(Z_{\star} - Z_{{\rm P}, \star}\right)\, X^{i}_{\rm T} + Z_{{\rm P}, \star}\, X^{i}_{\rm 60SN}
\end{aligned}
\end{equation}
Here, $Z^{i}_{\star}$ denotes the star particle's final mass fraction (not solar units) of $i \in$ [C, O, Mg, Ca, Fe] and where the subscripts $T$ and 60 SN denote the mass fractions of these elements in~\cite{Timmes} model and \cite{hegerwebsite}, respectively.
The results are displayed in Figure~\ref{fig:chem} in which we present mass-weighted, joint PDFs for all of the stars in our simulation. The probabilities have been determined using each star particle's corrected metallicity and primordial metallicity as described above. Once again we have used the upper bound for the correction to star particle metallicity. Using the lower bound does not change the probability distribution and hence does not make a difference in this analysis. The PDFs have also  been normalized by the logarithmic bin size and the comoving simulation volume.  As the \cite{Timmes} iron yields are [Fe/H] $\approx -0.5$, and very few $Z< 10^{-1.5} Z_\odot$ Pop II stars are formed below $z = 6,$ as shown in Figure \ref{fig:zpol}, we focus on the [Fe/H] range below -2, for which the $z=5$ results can be related to stars observed today.  In this range, for comparison, we have over-plotted the CEMP stars from \cite{2014Natur.506..463K}, with uncertainties, where known.

Without any fine tuning, our model results in reasonable agreement between the simulation's results and the observed characteristics of metal-poor stars, albeit using a single progenitor yield model.  Note that this is not guaranteed by our choice of Pop III yields, which were selected  by \cite{2014Natur.506..463K} to reproduce the properties of a single extremely metal-poor star, but are now being convolved with a wide range of $Z_{\star}$ and $Z_{{\rm P}, \star}$ values.   Nevertheless, our model captures many of the features observed in the data.  While [Fe/H] can reach extremely low values due the lack of iron in our PopIII yields, [C/H] and [O/H] are always greater than -4 and almost always greater than -3.
Note that the lack of such extremely carbon and oxygen poor stars  would not be reproduced in models that did not apply corrections to boost the observed metallicities of low  $\overline Z_{\star} $ star particles with small polluted fractions.  Furthermore, our simulations reproduce many other features suggested by the data including the enhanced probability of finding CEMP stars with [C/H] $\approx -1.5$ or with [C/Fe] $\approx 0.5,$ the enhanced probability of finding low-[Fe/H] stars with [O/H] $\approx -2,$ and the preponderance of low-[Fe/H] stars with  [Mg/Ca] $\approx 0. $

The histogram at the lower left of this figure depicts the mass-weighted metallicity density function (MDF) for our model overlaid with MDFs derived from MW halo populations \citep{2016arXiv160706336Y, 2013ApJ...763...65A}. To account for the fact that these have been derived from surveys targeted to find stars in a specific metallicity range, each of the observational MDFs has been normalized to the simulation's average MDF value over their respective intervals. Additionally, we have included 1 sigma Poisson noise based on the original data for each set of observations. 

Given the very low number of stars observed below [Fe/H] = -4.5 and since these MDFs were derived from local observational data over specific, and limited, metallicity ranges we do not expect a tight correlation with our global MDF at $z=5$. Specifically, the MDF for $-4.3 \leq$ [Fe/H] $ \leq -2.8$ exhibits selection effects based on the authors' focus on CEMP-no \citep{2016arXiv160706336Y} stars, not all metal poor stars and consists of high-resolution observations selected from medium resolution spectra likely to be extremely metal poor. On the other hand, the \citep{2013ApJ...763...65A} higher metallicity MDF,  [Fe/H] $>$ -3.8, is focused on medium-resolution spectroscopy that makes it hard to identify stars at the low end of the metallicity range, so we have chosen to use [Fe/H] $\geq -2.8$ for this data. 

\section{Conclusions}

Despite many recent advances in observations of MW halo stars with low metallicities, not a single star has yet been observed that does not contain some metals, pointing to the strong likelihood of a top-heavy Pop III IMF. Without direct observations to constrain the nature of such stars, it falls to theory to help us understand their properties and formation history. Yet modeling this evolution involves not only carrying out  large-scale cosmological simulations but also simultaneously tracking the enrichment of material on the much smaller scales in which stars form.

Put simply, the overriding factor that determines the transition from Pop III to Pop II star formation is the metallicity of the gas. While the critical metallicity for this transition is poorly constrained  \citep{2003Natur.422..869S, 2003Natur.425..812B, 2005ApJ...626..627O} it is expected to be $Z_{\rm crit} \approx 10^{-5}\,Z_{\odot},$  a level that is not reached until the ejecta from a single SNe are diluted into $\approx$ 10 million solar masses of material. This means that a key process affecting the formation of Pop II stars is the turbulent mixing of metals into pristine gas. On the other hand, when modeling a statistically significant number of early galaxies, it is too computationally expensive to model the composition of the gas down to star-forming scales. Thus, most works simply assume that metals injected into simulation cells instantaneously change the metallicity of the effected resolution elements, with subsequent star formation making use of these values. Others have used a high-resolution approach to study turbulent mixing within a single galaxy \citep{2010ApJ...716..510G}. 

In this work we have developed a new approach that allows us to track, statistically, the effects of subgrid turbulent mixing via a new scalar: the fraction of pristine gas, $P$, in each cell. We have used a self-convolution model to estimate  the rate at which turbulence  mixes pollutants thoroughly throughout a given volume of gas (PS 13), which is based only on two physical parameters: the average metallicity, $\overline Z$, of the pollutants/ejecta and the turbulent Mach number.  We have discovered that thorough mixing can take several eddy turnover times, demonstrating that modeling turbulent mixing is at least as important as modeling large-scale inhomogeneities when determining the Pop III star formation rate density. In fact, we find an increase of $2-3\times$ the Pop III SFR density as compared to similar cosmological simulations that do not account for subgrid mixing \citep{2014MNRAS.440.2498P, 2007MNRAS.382..945T}.  A natural follow-up to this work will be a parameter study in which we vary our stellar feedback prescriptions to quantify the effect on the SFR density.

As a result of modeling the pristine fraction of gas, we also improve our modeling of the metallicity of the polluted fraction, $f_{\rm pol}$, of gas and stars. Since the polluted gas fraction represents the vast majority of the metals in the cell that will eventually be mixed throughout its volume, we know that the metallicity of the fraction of polluted gas in the cell is $Z = \overline Z/f_{\rm pol}$.  By examining two representative halos, we find a significant difference between $Z_\star$ and $\overline Z_{\star}$   for stars with $\overline Z_{\star} < 10^{-3} Z_\odot,$ many of which have $f_{\rm pol} < 10^{-3}$. Assuming that Pop III/Pop II star formation in such incompletely mixed gas proceeds according to the fraction of polluted to pristine gas, we find that the effects of incomplete mixing are extremely important and help us to depict a more physical picture of metallicity evolution of early stars.

Given that the ejecta from the massive first stars is likely to be very different from that of subsequent stellar generations \citep{2002ApJ...567..532H}, we have also developed a straightforward method to track the metals generated by Pop III stars.  Our primordial metallicity scalar, $\overline Z_{\rm P}$ allows us to follow this material, and model the final metal content of stars as a convolution of two different types of material resulting in star particles with both $Z_{\rm P, \star}$ and $Z_{\star}$. This alleviates the need to model more detailed, and costly, chemical synthesis networks, and it allows us to quickly explore some of the parameter space suggested by others \citep{2014ApJ...792L..32I, 2003Natur.422..871U}, possibly ruling out certain SN progenitor ranges or yield models. In this work, we have modeled primordial metals as the result of $60 M_{\odot}$ Pop III SN \citep{2014Natur.506..463K, hegerwebsite}, demonstrating a possible origin for the metal levels seen in the sample of CEMP stars studied by \cite{2014Natur.506..463K}. In the future, we will explore a wider range of PopIII models,  hoping to provide insights for interpreting observations of the  chemical composition of low-metallicity stars in the local universe.

Finally, we plan to use our new tools to trace the evolution of and derive observational characteristics for a number of early galaxies. While the hunt for Pop III stellar populations is ongoing, and some promising candidates have been discovered \citep{2015ApJ...808..139S}, the upcoming James Webb Space Telescope and ground-based Giant Magellan Telescope, Thirty-meter Telescope, and European Extremely Large Telescope will open a much larger window on these objects. Simulations such as ours can help predict the spatial and temporal distribution of Pop III star formation needed to plan future searches for these objects.  We are still in the midst of developing models that capture the complex physics of PopIII star formation, and such tools will be crucial in defining the boundaries that help guide observers to find them. 

\acknowledgments

We would like to thank J. Devriendt, Y. Dubois, R. Teyssier, and M. Richardson for help understanding aspects of \textsc{ramses}. We would also like to thank T. Beers, A. Heger, F. X. Timmes, and J. Yoon  for helpful discussions.  This project was supported by NASA theory grant NNX15AK82G, and by the NSF under grant PHY 08-022648 for the Physics Frontier Center Joint Institute for Nuclear Astrophysics - Center for the Evolution of the Elements� (JINA-CEE). We  would like to thank the  Texas Advanced Computing  Center (TACC) at the University of Texas at Austin  and the Extreme  Science and Engineering Discovery Environment (XSEDE) for providing HPC resources via grant  TG-AST130021 that have contributed to the results reported within this paper. 

\clearpage 
\pagebreak

\end{document}